\documentclass[aps,prd,groupedaddress,showpacs,showkeys]{revtex4}  
\usepackage{latexsym,epsfig,amsmath,amssymb}  
\usepackage{amssymb}  
\usepackage{amsmath}  
\usepackage{amsfonts}  
\usepackage{epsfig}  
\usepackage{verbatim}  
\newcommand{\eq}{\begin{eqnarray}}  
\newcommand{\en}{\end{eqnarray}}

\begin{document}  
  
\title{  
On nature of the scalar-isoscalar mesons in the  
uniformizing-variable method \\  
based on analyticity and unitarity}  
  
\author{  
Yurii S. Surovtsev$^1$, Petr~Byd\v{z}ovsk\'y$^2$,  
Valery E. Lyubovitskij$^3$\footnote{On leave of  
absence from the Department of Physics, Tomsk State University,  
634050 Tomsk, Russia} \vspace*{1.2\baselineskip}} \affiliation{ $^1$  
Bogoliubov Laboratory of Theoretical Physics, Joint  
Institute for Nuclear Research, 141 980 Dubna, Russia\\  
\vspace*{.4\baselineskip}\\  
$^2$ Nuclear Physics Institute, Czech Academy of Sciences, \v{R}e\v{z}  
near Prague 25068, Czech Republic\\  
\vspace*{.4\baselineskip}\\  
$^3$ Institut f\"ur Theoretische Physik,  Universit\"at T\"ubingen,\\  
Kepler Center for Astro and Particle Physics, \\  
Auf der Morgenstelle 14, D--72076 T\"ubingen, Germany\\}

\date{\today}  
  
\begin{abstract}  
  
The experimental data on the processes  
$\pi\pi\to\pi\pi,K\overline{K},\eta\eta,\eta\eta^\prime$ in the  
$I^GJ^{PC}=0^+0^{++}$ channel have been jointly analyzed to  
study the status and nature of the $f_0$. The method of analysis  
is based on analyticity and unitarity and uses an uniformization  
procedure. Some spectroscopic implications from results of  
the analysis are discussed.  
  
\end{abstract}  
  
\pacs{11.55.Bq,11.80.Gw,12.39.Mk,14.40.Cs}  
  
\keywords{coupled--channel formalism, meson--meson scattering,  
scalar and pseudoscalar mesons}  
  
\maketitle  
  
\section{Introduction}  
  
The problem of interpretation of scalar mesons is tightly related  
to the most profound topics in particle physics which concern the  
QCD vacuum (see, e.g., the review ``Note on scalar mesons''  
in~\cite{PDG10}). These mesons are expected to be composed of the  
$q{\bar q}$ or the lightest 4-quark states, meson-meson molecules  
or gluonium states. It is disconcerning that up to now a description  
of this mesonic sector is far from being complete despite of the big  
effort devoted to studying various aspects of the problem (for recent  
reviews see, e.g.~\cite{Amsler04,Bugg04,Close02,Klempt07}). Parameters  
of the scalar mesons, their nature and status of some of them is still  
not settled~\cite{PDG10}. Especially, this concerns the  
$f_0(600)/\sigma$ meson and $K_0^*(900)/\kappa(800)$ meson.  
For example, the mass of the former obtained in the Breit--Wigner  
or $K$-matrix approaches ranges in various analyses in the interval  
of about 400--1200~MeV \cite{PDG10}. According to the prediction  
by Weinberg~\cite{Wei90} based on the mended symmetry the mass  
of the $\sigma$ should be near the mass of the $\rho$-meson. As to the  
mass of the lowest scalar glueball, various non-perturbative QCD methods  
give also very different results. From the QCD sum rules~\cite{KamMN09}  
one has found a scalar-isoscalar meson of the gluonium nature with  
a mass about 1000~MeV and with the $\pi\pi$-decay width about 500~MeV.  
This is in agreement with the recent unquenched-lattice simulation  
using dynamical fermions \cite{McNeile09} but it diverges from recent  
calculations on the quenched anisotropic lattices of the glueball  
spectrum where the mass of the lowest glueball is about  
1710~MeV~\cite{Chen06}.  
  
The width of the $f_0(600)$ (in various experiments and analyses) also  
has a large spread 600--1000~MeV according to an estimate of the  
Particle Data Group team \cite{PDG10}. Note also the works  
in which one obtained a very small value of $35\pm 12$~MeV~\cite{Troyan}  
and the very large one of  about 3200~MeV \cite{Achasov-Shest}.  
The prediction for the $\sigma$-meson width  
on the basis of saturating the superconvergence dispersive sum rules  
is larger than about 670 MeV~\cite{Gilman}. The theoretical  
conclusions about widths of glueballs, especially about the lightest one,  
are also very different in various approaches. In  
Ref.~\cite{Ellis-Lanik} the authors used an effective QCD Lagrangian  
with the broken scale and chiral symmetry, where a glueball is  
introduced to theory as a dilaton and its existence is related to  
breaking of scale symmetry in QCD. Then the $\pi\pi$ decay width of  
the glueball, estimated using low-energy theorems, is $\Gamma(G \to  
\pi\pi)\approx 0.6\,{\rm GeV}\times(m_G/1\,{\rm GeV})^5$, where  
$m_G$ is the glueball mass. I.e., for the glueball with the mass  
about 1~GeV (if it exists), the width is near 600 MeV. Though  
a use of the above formula is doubtful above 1~GeV, a tendency for  
the glueball to be wide is apparently seen. This is supported by  
arguments given in \cite{Anisovich98} that the glueball width  
is larger than the ones of the surrounding $q{\bar q}$ states. On  
the other hand, in Ref.~\cite{GGLF}, where the two-pseudoscalar and  
two-photon decays of the scalars between 1--2~GeV were analyzed in  
the framework of a chiral Lagrangian and the glueball was included  
as a flavor-blind composite mesonic field, the glueball was found  
to be rather narrow in accordance with the  
former findings of Ref.~\cite{Ams96}.  
  
Up to now the nature of the $f_0(980)$ is not clearly resolved.  
Besides a $q{\bar q}$ \cite{Morgan,Torn,Lanik,Tornqvist2}, subject  
to serious criticism, there are recent arguments for a 4-quark  
state (as the ${a_0}(980)$)~\cite{Achasov00}, a $K\overline{K}$  
molecule \cite{Isgur,Jansen,BGL} and a $\eta\eta$ bound state  
\cite{Sur-Kam_07,SBKN-ijmp09,SBKN-prd10}.  
  
Existence of the $f_0(1370)$ meson is still not obvious.  
In some works, e.g., in \cite{MO-02-2,Ochs10} one did not find  
any evidence for the existence of the $f_0(1370)$. In  
Ref.~\cite{SKN-epja} also the best description of  
$\pi\pi\to\pi\pi,K\overline{K}$ was obtained without the  
$f_0(1370)$, and it was shown that the $K\overline{K}$ scattering  
length is very sensitive to whether this state exists or not. On the  
other hand, in Ref.~\cite{Bugg1370} a number of data requiring  
apparently the existence of the $f_0(1370)$ is indicated: the  
Crystal Barrel data on ${\bar p}p\to\eta\eta\pi^0$ \cite{Amsler92}  
and on ${\bar p}p\to3\pi^0$ \cite{Abele96}, the BES data on  
$J/\psi\to\phi\pi^+\pi^-$ \cite{Ablikim05}; the $f_0(1370)$ appears  
also in the GAMS data for $\pi^+\pi^-\to\pi^0\pi^0$ at large $|t|$  
\cite{Alde98}. For example, in \cite{Oset1370} it was shown within  
the so-called ``hidden gauge formalism'' that the $f_0(1370)$ might be  
dynamically generated from the $\rho\rho$ interaction.  
  
Especially it is worth to discuss the situation with scalar states in  
the 1500~MeV region. First, a state, observed in this region could be  
a real candidate for the lightest glueball (see, e.g., Ref.~\cite{Ams96}).  
In the model-independent analyses of data on the processes  
$\pi\pi\to\pi\pi\,K\overline{K},\eta\eta,\eta\eta^{\prime}$ using  
different uniformizing variables~\cite{Sur-Kam_07,SBKN-ijmp09,SBKN-prd10,  
SKN-epja,PRD-01,SKN-AIP04,SKN-czjp06,KMS96}, a wide state  
$f_0(1500)$ was obtained whereas in many other works, which analyzed  
mainly the production and decay of mesons, as cited in the PDG tables  
\cite{PDG10}, the rather narrow $f_0(1500)$ is obtained.  
Therefore, we have supposed \cite{SKN-epja,SKN-AIP04} that the wide  
$f_0(1500)$, observed in the multi-channel $\pi\pi$ scattering,  
indeed, is a superposition of two states, narrow ($q{\bar q}$) and  
broad (glueball). The former is just observed in the processes of decay  
and production of mesons. An indication about nature of the latter  
follows from the fact that the $f_0(1500)$ is coupled with the  
approximately equal strength with the $\pi\pi$, $K\overline{K}$ and  
$\eta\eta$ systems  
\cite{Sur-Kam_07,SBKN-ijmp09,SKN-epja,SKN-AIP04,PRD-01,SKN-czjp06}  
and from the arguments of Ref.~\cite{Anisovich98} on the glueball  
width. These suppositions are in some accordance with the results of  
the combined K-matrix analysis \cite{Anisovich02} of the GAMS data  
on $\pi^-p\to\pi^0\pi^0n,\eta\eta n,\eta\eta^{\prime}n$ \cite{Alde95},  
BNL data on $\pi^-p\to K\overline{K}n$~\cite{Lindenbaum92} and Crystal  
Barrel data on $p{\bar p}\,(at~rest)\to\pi^0\pi^0\pi^0,  
\pi^0\pi^0\eta,\pi^0\eta\eta$ \cite{Anisovich94,Amsler95}, which say  
that in the 1500~MeV region there are the narrow $f_0(1500)$ and  
very wide $f_0(1530^{+90}_{-250})$.  
  
The $f_0(1710)$ has most likely the dominant $s{\bar s}$ component  
(see, e.g., Refs.~\cite{SKN-epja,Close-Kirk2} and the review ``Note  
on scalar mesons'' in~\cite{PDG10}). Note, however, that the QCD sum  
rules \cite{Narison} and the $K$-matrix method \cite{Anisovich}  
showed that both $f_0(1500)$ and $f_0(1710)$ are mixed states with  
a large admixture of the glueball component. There are also schemes  
\cite{MO-02-2,Lee-Weing00} in which the coupling of two gluons (of a  
scalar glueball) with $n{\bar n}$ ($n$ is nonstrange $u$ or $d$ quark)  
appears to be suppressed by chiral symmetry \cite{Chanowitz} increasing the  
relative contribution of the $s{\bar s}$ component. When assuming  
this consideration to be valid up to energies of the $f_0(1710)$,  
one concludes that this state could be an unmixed glueball  
\cite{Alb-Oller}.  
  
In the scalar-isodoublet sector (except for the well-established state  
$K_0^*(1430)$) the possible existence of a very broad meson in the  
700-950~MeV region is discussed in recent years (see the review ``Note  
on scalar mesons'' in~\cite{PDG10}). E.g., in some recent analyses  
the authors have found a pole which corresponds to this state  
$K_0^*(900)$  
\cite{Bugg06,Ablikim06C,Descotes-Genon06,Zhou06,Cawlfield06A,Link07B,  
SBGL-ppn10}, while no such state was seen in the experiment performed  
by the BaBar Collaboration~\cite{Aubert07T} and in the earlier  
analyses \cite{Che-Penn01,Kopp01,Link05}.  
  
In view of all above circumstances, the problems connected with  
determining the nature of the observed mesonic states and their  
assignment to the quark-model configurations are still open in  
spite of a large amount of work devoted to these problems (see,  
e.g., Refs.~\cite{GI,Tornqvist,An-Prok-Sar,Close-Torn,Volkov-Yudich}).  
It is clear that resonance parameters should be obtained, if  
possible, in a model-independent way. Here, we present results  
of the combined three-channel analysis of data on the processes  
$\pi\pi\to\pi\pi,K\overline{K},\eta\eta,\eta\eta^\prime$ in the  
channel with the quantum numbers $I^GJ^{PC}=0^+0^{++}$. Study of  
the $K\pi$ scattering in the channel with $I(J^P)=\frac{1}{2}(0^{+})$  
and the role of the strange scalar meson $K_0^*(900)$ ($\kappa(800)$)  
goes beyond the scope of this paper and will be discussed in  
Ref.~\cite{SBL}. We have used a ``model-independent'' method  
\cite{SBKN-prd10,SKN-epja,PRD-01,SKN-czjp06,KMS96} based on the  
first principles (analyticity and unitarity) directly applied to  
the analysis of experimental data. This approach permits us to omit  
a theoretical prejudice in extracting the resonance parameters. It  
is important that an uniformizing-variable method allows to avoid  
a model dependence when considering resonance  
contributions. This is possible since a main model-independent contribution  
of the resonance can be given by poles and corresponding zeros  
on a uniformization plane, whereas the  
possible remaining corrected and model-dependent contribution of the  
resonance is supposed to be taken into account in the background.  
This distinguishes substantially our model-independent method from  
the standard dispersion relation approach based also on analyticity  
and unitarity where, however, the model dependence arises inevitably  
when saturating dispersive integrals by the contributions of  
resonances. Then in our method, considering the obtained disposition 
of resonance  
poles on the Riemann surface, bearing witness to a relative strength  
of coupling with corresponding channels, and resonance masses,  
we draw conclusions about nature of the investigated states.  
  
Unlike in the previous three-channel analysis of the above  
processes \cite{Sur-Kam_07,SBKN-ijmp09,SBKN-prd10,SKN-czjp06}, in this  
work we used a new uniformizing variable in which   
we took into account the left-hand branch-point at  
$s=0$ related to the thresholds of the $\pi\pi$  
scattering in crossed channels, in addition to the right-hand  
branch-points related to the thresholds of the analyzed processes.  
This should diminish considerably dependence of the extracted  
parameters of resonances on the background because the elastic part  
of the $\pi\pi$ background is stipulated mainly by the contribution  
of the left-hand cuts.  
  
The layout of the paper is as follows. In Sec. II we outline the two-  
and three-coupled channel formalism, where pole clusters on the Riemann  
surface are determined as characteristics of multichannel states and  
a classification of two- and three-channel resonances according to the  
types of the possible pole-clusters is given. We introduce also the    
new uniformizing variable for the three-channel case  
taking into account the left-hand branch-point at $s=0$, and show the  
disposition of the resonance poles and zeroes related to the various  
pole-clusters on the unifomization plane for the $\pi\pi$-scattering  
$S$-matrix element. In Sec. III we carry out the combined 3-channel  
analyses of data on the processes $\pi\pi\to\pi\pi,K\overline{K},\eta\eta$  
(variant I) and $\pi\pi\to\pi\pi,K\overline{K},\eta\eta^\prime$ (variant  
II). In Sec. IV we summarize our conclusions,  
propose an assignment of the scalar mesons lying below 1.9 GeV to lower  
nonets and discuss the obtained results.  
  
\section{The coupled-channel formalism in model-independent approach }  
  
Our model-independent method which utilizes an uniformizing variable  
can be used only for the 2-channel case and under some conditions for  
the 3-channel one. Only in these cases we obtain a simple symmetric  
(easily to be interpreted) picture of the resonance poles and zeros of  
the $S$-matrix on the uniformization plane. The $S$-matrix is determined  
on the 4- and 8-sheeted Riemann surfaces for the 2- and 3-channel cases,  
respectively. The matrix elements $S_{\alpha\beta}$, where  
$\alpha,\beta=1,2,3$ denote the channels, have the right-hand cuts along  
the real axis of the complex $s$ plane ($s$ is the invariant total  
energy squared), starting with the channel thresholds $s_i$ ($i=1,2,3$),  
and the left-hand cuts related to crossed channels. The Riemann-surface  
sheets are numbered according to the signs of the analytic continuations  
of the quantities $\sqrt{s-s_\alpha}$ as follows: in the 2-channel case  
\eq\mbox{signs}\Big(\mbox{Im}\sqrt{s-s_1},\mbox{Im}\sqrt{s-s_2}\Big)=  
++,-+,--,+- \en correspond to sheets I, II, III, IV; in the  
3-channel case \eq \mbox{signs}\Big(\mbox{Im}\sqrt{s-s_1},  
\mbox{Im}\sqrt{s-s_2},{\mbox{Im}}\sqrt{s-s_3}\Big)=  
+++,-++,--+,+-+,+--,---,-+-, ++- \en correspond to sheets I,  
II,$\cdots$, VIII, respectively.  
  
The resonance representations on the Riemann surfaces are obtained using  
formulas from Ref.~\cite{KMS96} (see also Appendix), expressing  
analytic continuations of the $S$-matrix elements to unphysical sheets  
in terms of those on sheet I (physical) that  
have only the resonance zeros (beyond the real axis)  
around the physical region. Then, starting from the resonance  
zeros on sheet I one can obtain an arrangement of poles and zeros of  
resonances on the whole Riemann surface.  
  
In the 2-channel case we obtain tree types of resonances described by  
a pair of conjugate zeros on sheet I only in $S_{11}$ -- the type  
({\bf a}), only in $S_{22}$ -- ({\bf b}), and in each of $S_{11}$  
and $S_{22}$ -- ({\bf c}). Then the formulas of the analytic  
continuations of the $S$-matrix elements to unphysical  
sheets~\cite{KMS96} (see also Appendix) immediately give the  
resonance representation by poles and zeros on the 4-sheeted Riemann  
surface: to the resonances of types ({\bf a}) and  ({\bf b}), there  
corresponds a pair of complex conjugate poles on sheet III shifted  
relative to a pair of poles on sheet II and IV, respectively. For  
the states of type ({\bf c}) one must consider the corresponding  
two pairs of conjugate poles on sheet III.  
  
In the 3-channel case we obtain seven types of resonances corresponding  
to seven possible situations when there are resonance zeros on sheet I  
only in $S_{11}$ -- ({\bf a}); ~~$S_{22}$ -- ({\bf b}); ~~$S_{33}$  
-- ({\bf c}); ~~$S_{11}$ and $S_{22}$ -- ({\bf d}); ~~$S_{22}$ and  
$S_{33}$ -- ({\bf e}); ~~$S_{11}$ and $S_{33}$ -- ({\bf f});  
~~$S_{11}$, $S_{22}$, and $S_{33}$ -- ({\bf g}). Examples for the  
disposition of poles and zeros, corresponding to some of these types  
of the 3-channel resonances, on the unifomization plane will be  
given in the next section.  
  
The resonance of every type is represented by a pair of  
complex-conjugate clusters (of poles and zeros on the Riemann  
surface). The cluster type is related to the nature of the state. For  
example, if we consider the $\pi\pi$, $K\overline{K}$ and $\eta\eta$  
channels, a resonance coupled relatively more strongly to the  
$\pi\pi$ channel than to the $K\overline{K}$ and $\eta\eta$ ones is  
described by the cluster of type ({\bf a}). In the opposite case, it  
is represented by the cluster of type ({\bf e}) (the state with a  
dominant $s{\bar s}$ component). The glueball must be represented by  
the cluster of type ({\bf g}) (of type ({\bf c}) in the 2-channel  
consideration) as a necessary condition for the ideal case, if this  
state lies above the thresholds of the considered channels.  
  
A necessary and sufficient condition for existence of the  
multi-channel resonance is its representation by one of the types of  
pole clusters. Note that whereas cases ({\bf a}), ({\bf b}) and  
({\bf c}) can be simply related to the representation of resonances  
by multi-channel Breit--Wigner forms, cases ({\bf d}), ({\bf e}),  
({\bf f}) and ({\bf g}) are practically lost in the Breit--Wigner  
description.  
  
One can formulate a model-independent test as a necessary condition  
to distinguish a bound state of colorless particles ({\it e.g.}, a  
$K\overline{K}$ molecule) and a $q{\bar q}$ bound state  
\cite{PRD-01,KMS96,MP93}. In the 1-channel case, the  
existence of the particle bound-state corresponds to the presence of a pole  
on the real axis under the threshold on the physical sheet. In  
the 2-channel case the existence of the bound-state in channel 2  
($K\overline{K}$ molecule) that, however, can decay into channel 1  
($\pi\pi$ decay), would imply the presence of the pair of complex  
conjugate poles on sheet II under the second-channel threshold  
without the corresponding shifted pair of poles on sheet III.  
  
In the 3-channel case the bound state in channel 3 ($\eta\eta$)  
that can decay into channels 1 ($\pi\pi$ decay) and 2  
($K\overline{K}$ decay) is represented by the pair of complex  
conjugate poles on sheet II and by the pair of shifted poles on  
sheet III under the $\eta\eta$ threshold without the corresponding  
poles on sheets VI and VII. According to this test, an interpretation  
of the $f_0(980)$ as the $K\overline{K}$ molecule was rejected earlier  
\cite{KMS96}. The reason is that this state is  
represented by the cluster of type ({\bf a}) in the 2-channel  
analysis of processes $\pi\pi\to\pi\pi,K\overline{K}$ and,  
therefore, does not satisfy the necessary condition to be the  
$K\overline{K}$ molecule. A further discussion of this topic will be  
given in the last section.  
  
Unlike the standard dispersion relation approach in our  
model-independent method we use an advantage of the fact that the  
amplitude is a one-valued function on the Riemann surface. To this  
end a uniformizing variable is applied, which maps the Riemann  
surface onto a complex plane. This permit us to use the representation by  
the pole clusters very important for the broad multi-channel resonances.  
This is impossible in the standard dispersion relation and  
$K$-matrix approaches or in the Breit--Wigner one.  
  
In the combined analysis of coupled processes it is convenient to  
use the Le Couteur--Newton relations~\cite{LeCou}. They express the  
$S$-matrix elements of all coupled processes in terms of the Jost  
matrix determinant $d(\sqrt{s-s_1},\cdots,\sqrt{s-s_N})$ that is a  
real analytic function with the only square-root branch-points at  
$\sqrt{s-s_\alpha}=0$. The important branch points, corresponding to  
the thresholds of the coupled channels and to the crossing ones, are  
taken into account in the proper uniformizing variable. On the  
uniformization plane, the pole-cluster representation of the  
resonance is the good one.  
  
It is obvious that the main model-independent contribution of  
resonances is factorized in the $S$-matrix elements from the  
background. The possible remaining corrected and model-dependent  
contributions of resonances are supposed to be included in the background.  
This is realized in a natural way: in the background, the  
corresponding elastic and inelastic phase shifts increase when  
some channel is opened. Therefore, we denote our approach as model  
independent.  
  
In the 2-channel case, the $S$-matrix, determined on the 4-sheeted  
Riemann surface, can be uniformized using, e.g., the inverse  
Zhukovskij transformation~\cite{KMS96} in which the thresholds of  
two channels are taken into account. Unfortunately, already in the  
3-channel consideration a  
function determined on the 8-sheeted Riemann surface can be  
uniformized only on torus. This is unsatisfactory for our purposes.  
Therefore, we neglect the influence of the lowest ($\pi\pi$)  
threshold branch-point at $s_1$ (however, unitarity on the  
$\pi\pi$-cut is taken into account). An approximation like this  
means the consideration of the nearest to the physical region  
semi-sheets of the Riemann surface of the $S$-matrix. In fact, we  
construct a 4-sheeted model of the initial 8-sheeted Riemann surface  
approximating it in accordance with our approach of a consistent  
account of the nearest singularities on all the relevant sheets. In  
practice, neglecting the influence of the $\pi\pi$-threshold  
branch-point means that we do not describe some small region near  
the threshold. Furthermore, we shall take into account in the  
uniformizing variable also the left-hand branch-point at $s=0$  
related to the crossed $\pi\pi$ channels. The allowance for this branch-point  
should diminish the background dependence of the obtained results.  
As was indicated repeatedly (see, e.g., Ref.~\cite{Achasov-Shest}) many  
analyses are subject to criticism (especially with the point of view of proof  
of the resonance existence) because the wide-resonance parameters  
are strongly controlled by the non-resonant background; this is particularly  
related to low-lying states. The allowance for the left-hand  
branch-point, related to the crossed channels, serves in part for a  
solution of this problem. For example in Ref.~\cite{SKN-epja} a  
combined analysis of the processes $\pi\pi\to\pi\pi,K\overline{K}$  
in the isoscalar-scalar sector was performed using the method of a  
uniformizing variable  
which includes two threshold branch-points and the left-hand one at  
$s=0$. In this respect a parameterless description of the $\pi\pi$  
background was obtained due to the inclusion of the indicated left-hand  
branch-point. Moreover, it was shown that the large  
background, obtained in earlier analyses of the $S$-wave $\pi\pi$  
scattering~\cite{KMS96}, hides in reality the $\sigma$-meson lying  
below 1~GeV.  
  
\begin{figure}[htb]  
\begin{center}  
\includegraphics[width=0.45\textwidth,angle=-90]{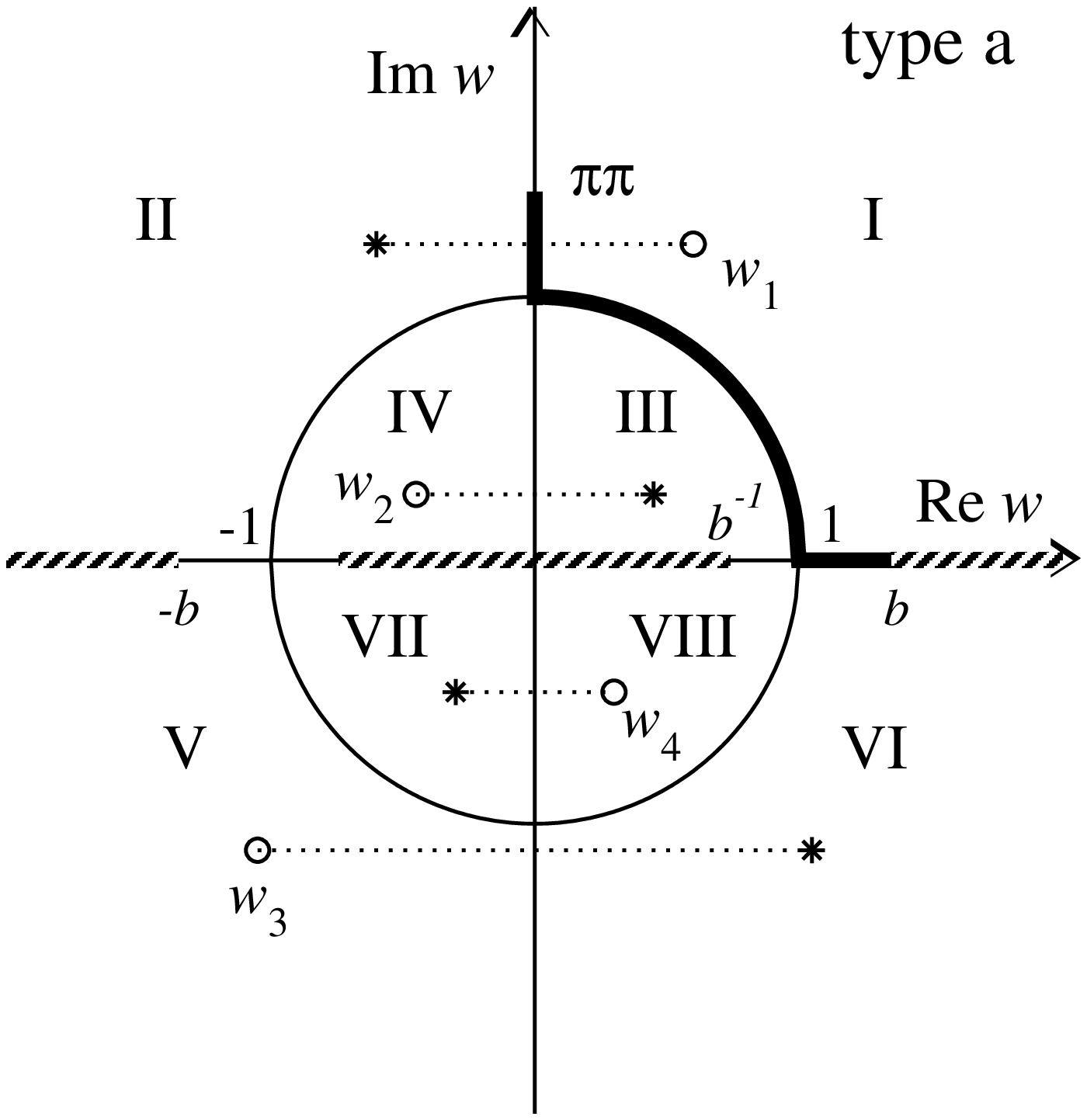}  
\includegraphics[width=0.45\textwidth,angle=-90]{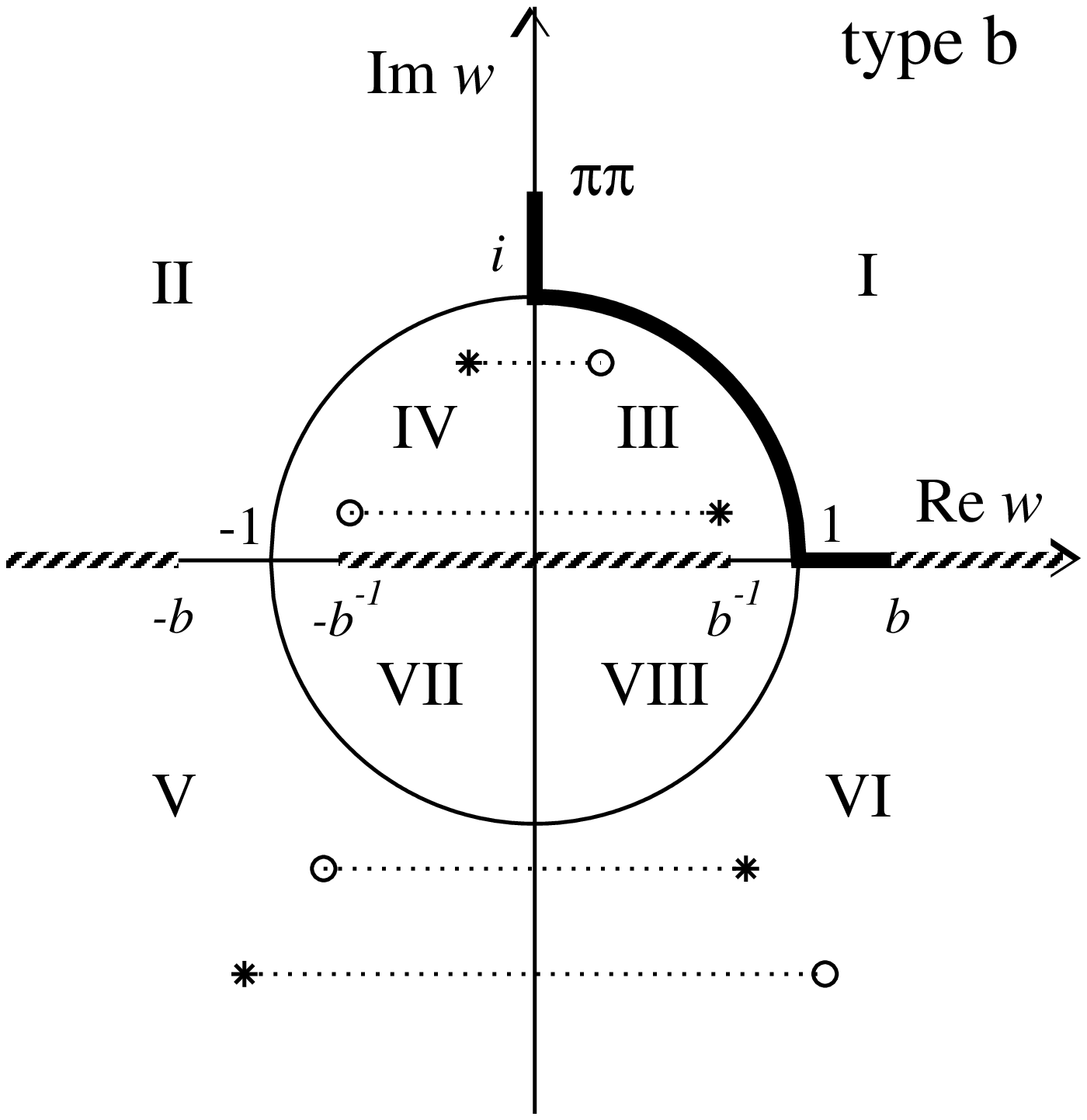}\\  
\includegraphics[width=0.45\textwidth,angle=-90]{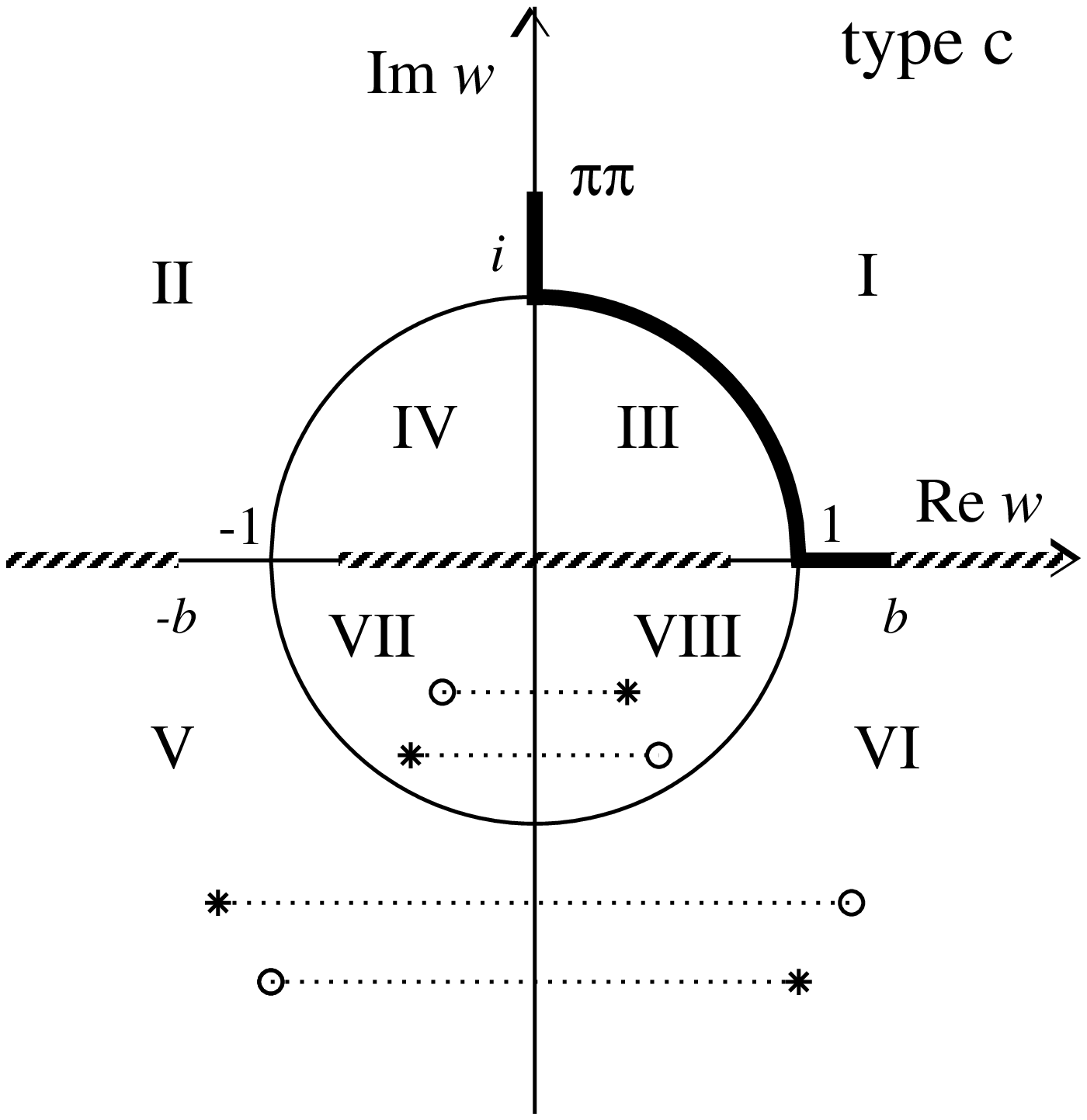}  
\includegraphics[width=0.45\textwidth,angle=-90]{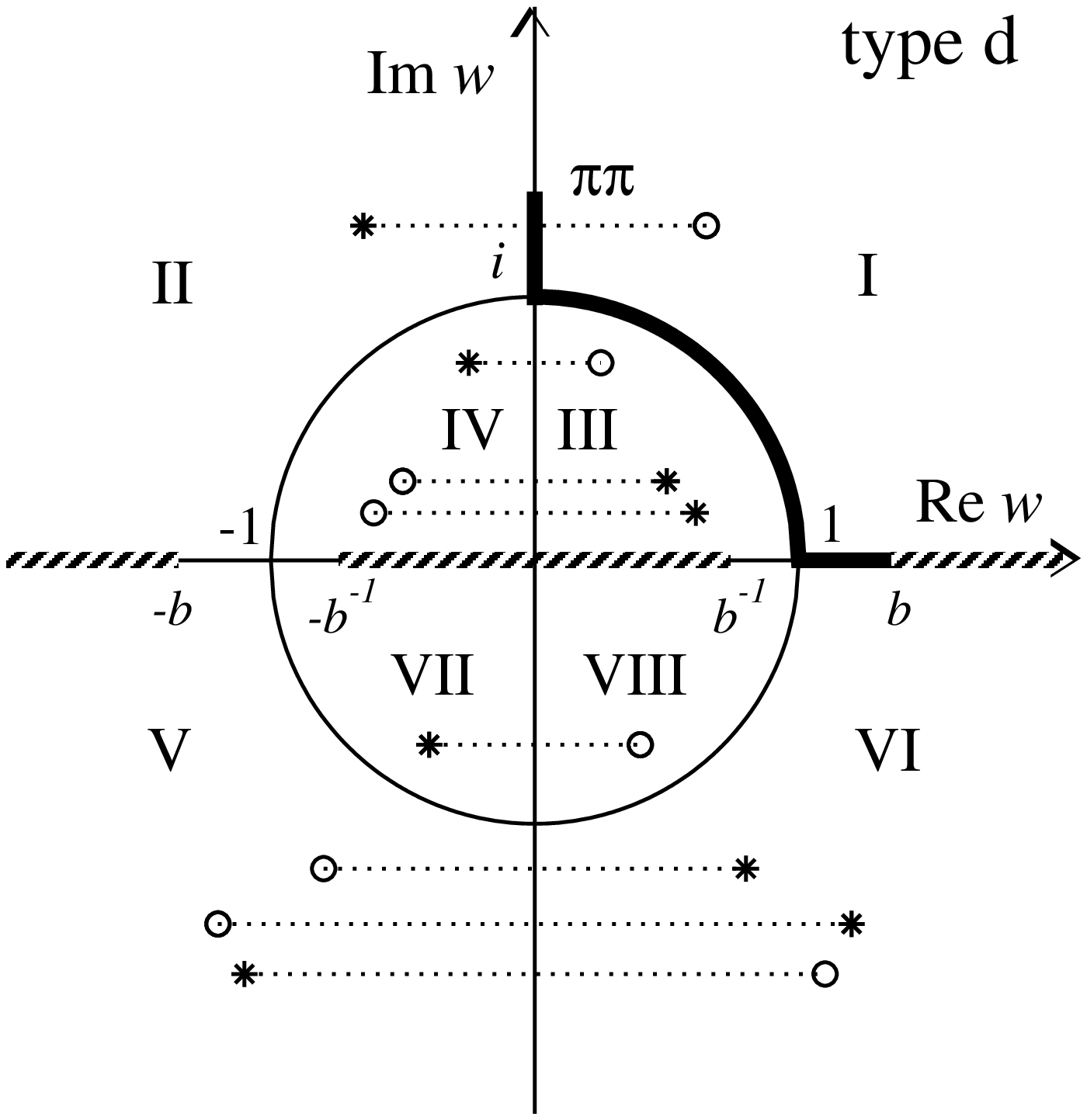}  
\caption{Uniformization $w$-plane for the 3-channel-$\pi\pi$-scattering  
amplitude. Representation of resonances of types ({\bf a}), ({\bf b}),  
({\bf c}) and ({\bf d}) is shown.}  
\label{fig:lw_plane_abcd}  
\end{center}  
\end{figure}

\begin{figure}[htb]  
\begin{center}  
\includegraphics[width=0.45\textwidth,angle=-90]{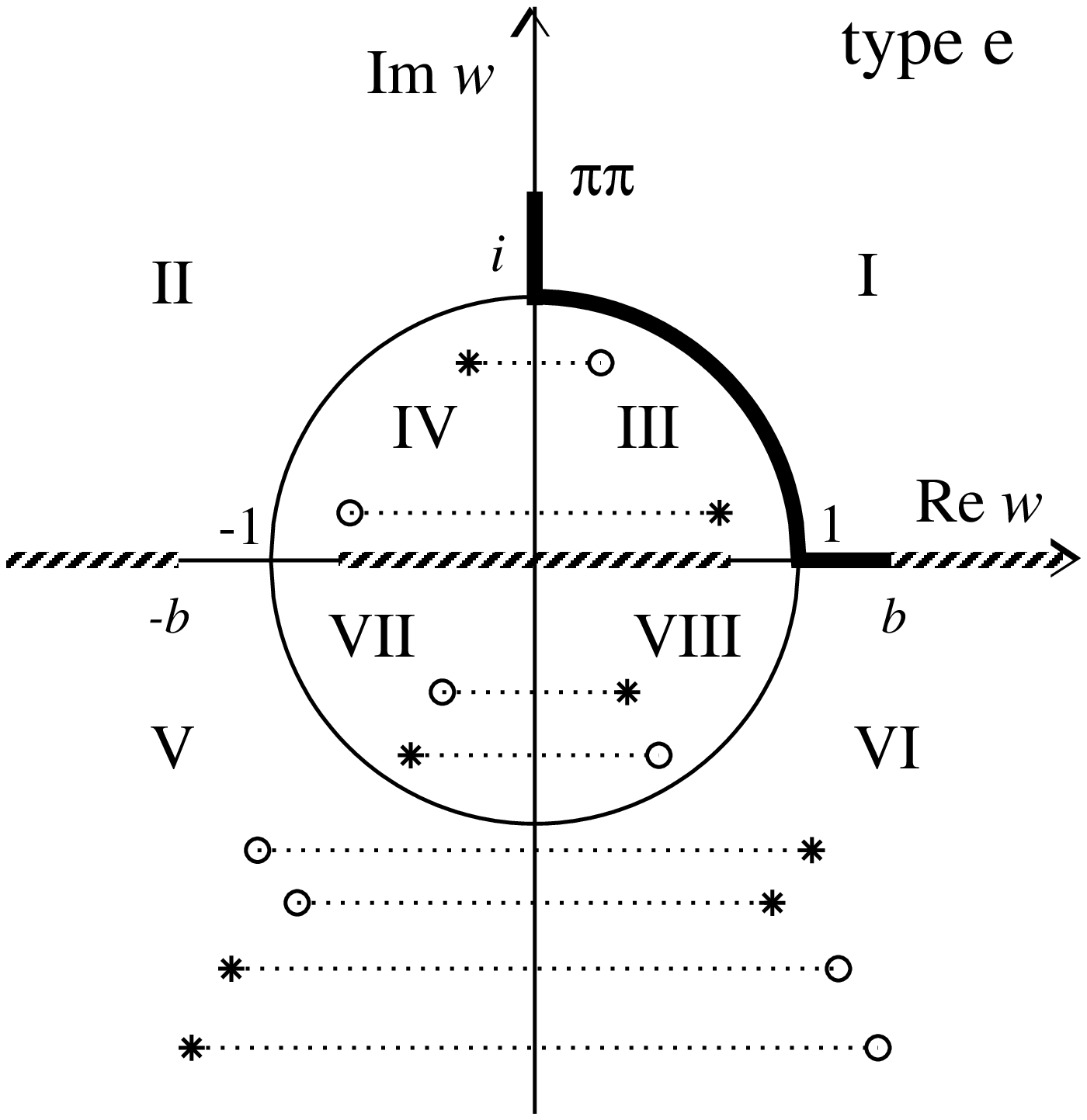}  
\includegraphics[width=0.45\textwidth,angle=-90]{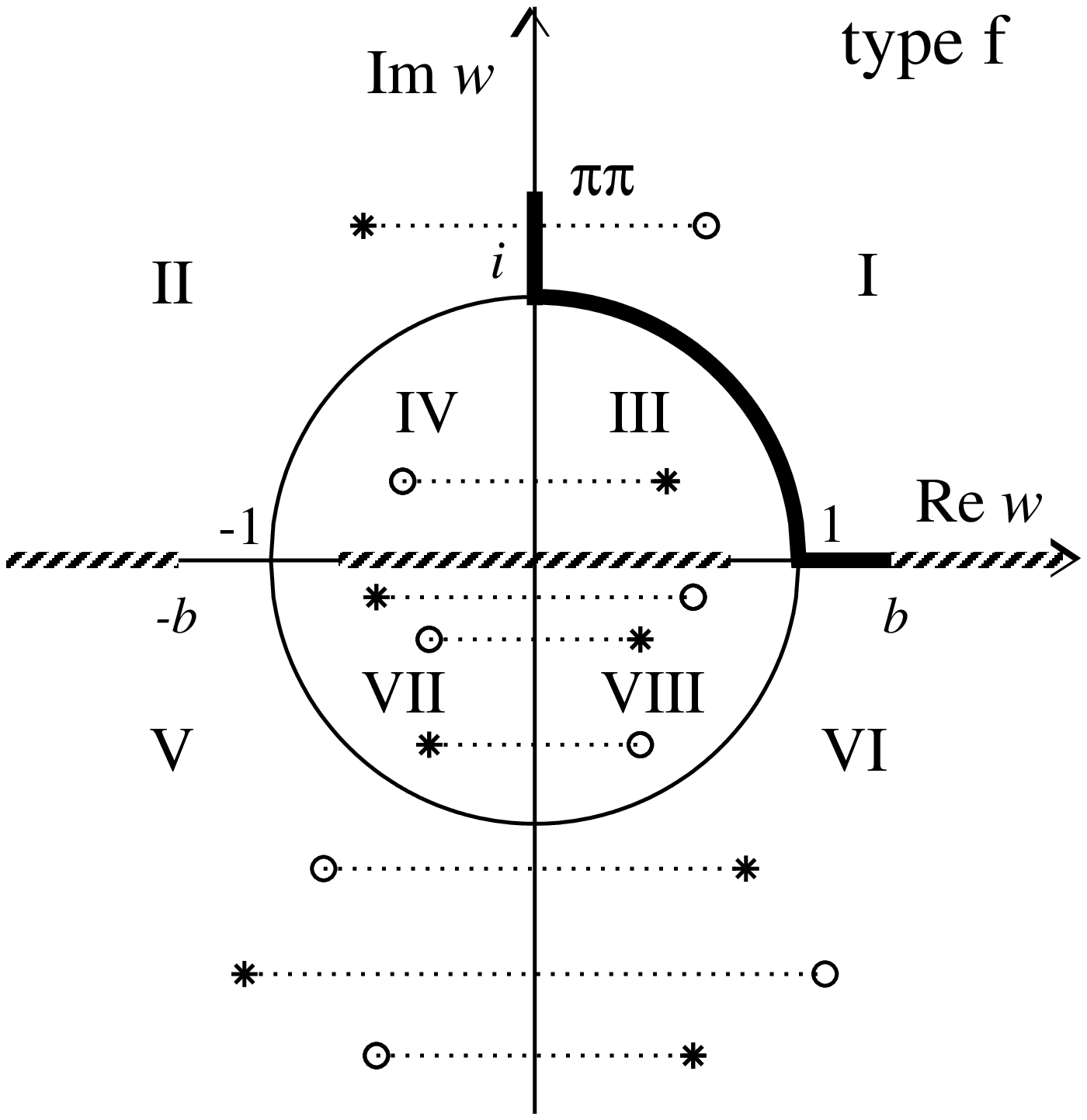}\\  
\includegraphics[width=0.45\textwidth,angle=-90]{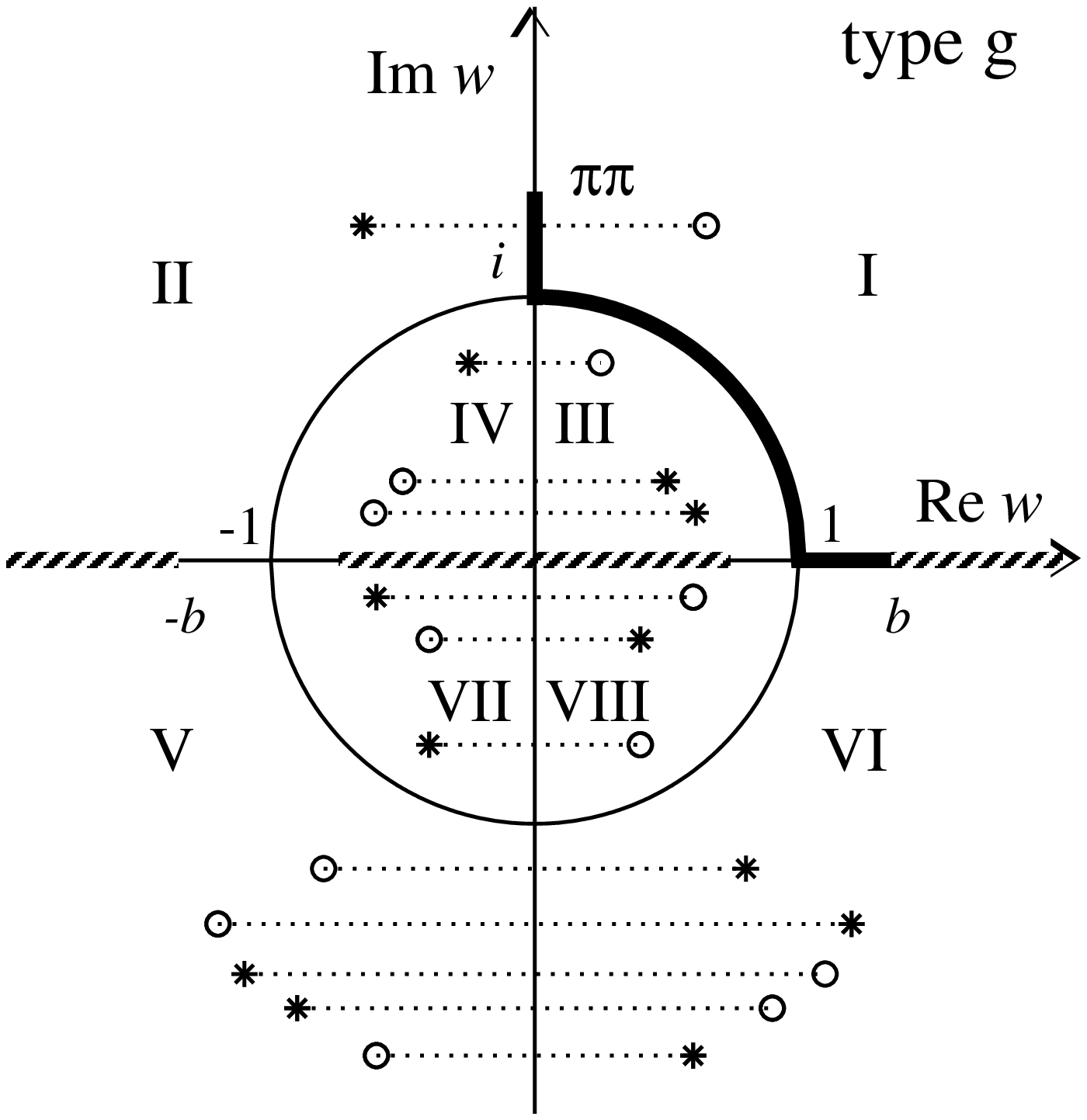}  
\caption{Uniformization $w$-plane for the 3-channel scattering  
amplitude. Representation of resonances of types ({\bf e}), ({\bf f}),  
and ({\bf g}) is shown.}  
\label{fig:lw_plane_efg}  
\end{center}  
\end{figure}

In the 3-channel case the new uniformizing variable used below  
can have the form:  
\begin{equation}\label{w}  
w=\frac{\sqrt{(s-s_2)s_3} + \sqrt{(s-s_3)s_2}}{\sqrt{s(s_3-s_2)}}  
\end{equation}  
where we neglect the lowest $\pi\pi$-threshold branch-point and  
take into account the threshold branch-points related to two  
remaining channels and the left-hand branch-point at $s=0$.  
This variable maps our model of the 8-sheeted Riemann surface onto the  
$w$-plane divided into two parts by a unit circle centered at the  
origin. The semi-sheets I (III), II (IV), V (VII) and VI (VIII) are  
mapped onto the exterior (interior) of the  
unit disk in the 1st, 2nd, 3rd and 4th quadrants, respectively. The  
physical region extends from the point $\pi\pi$ on the imaginary  
axis (the first $\pi\pi$ threshold corresponding to $s_1$) along this  
axis down to the point {\it i} on the unit circle (the second threshold  
corresponding to $s_2$). Then it extends further along the unit circle  
clockwise in the  
1st quadrant to point 1 on the real axis (the third threshold  
corresponding to $s_3$) and then along the real axis to the point  
$b=(\sqrt{s_2}+\sqrt{s_3})/\sqrt{s_3-s_2}$ into which $s=\infty$ is  
mapped on the $w$-plane. The intervals  
$(-\infty,-b]$, $[-b^{-1},b^{-1}]$, $[b,\infty)$ on the real axis are  
the images of the corresponding edges of the left-hand cut of the  
$\pi\pi$-scattering amplitude. In Figs.\ref{fig:lw_plane_abcd} and  
\ref{fig:lw_plane_efg}, the 3-channel resonances of all the standard  
types in $S_{11}(w)$ are represented  by the poles ($*$) and zeroes  
($\circ$) symmetric to these poles with respect to the imaginary axis  
giving corresponding pole clusters. The ``pole--zero'' symmetry  
guarantees the elastic unitarity of $\pi\pi$ scattering in the  
($\pi\pi$, $i$) interval.  
%
%
%
%
For the data we use the results of phase analyses which are  
given for phase shifts of the amplitudes $\delta_{\alpha\beta}$ and  
for the modules of the $S$-matrix elements  
$\eta_{\alpha\beta}=|S_{\alpha\beta}|$ ($\alpha,\beta=1,~2,~3$):  
\begin{equation}  
S_{\alpha\alpha}=\eta_{\alpha\alpha}e^{2i\delta_{\alpha\alpha}},~~~~~  
S_{\alpha\beta}=\eta_{\alpha\beta}e^{i\phi_{\alpha\beta}}.  
\end{equation}  
If below the third threshold there is the 2-channel unitarity then  
the relations  
\begin{equation}  
\eta_{11}=\eta_{22}, ~~ \eta_{12}=(1-{\eta_{11}}^2)^{1/2},~~  
\phi_{12}=\delta_{11}+\delta_{22}  
\end{equation}  
are fulfilled in this energy region.  
  
Masses and total widths can be calculated using the denominator of  
formula for the resonance part of the amplitude of the form  
\begin{equation} \label{T_res}  
T^{res}=\frac{\sqrt{s}\Gamma_{el}}{m_{res}^2-s-i\sqrt{s}\Gamma_{tot}}  
\end{equation}  
taking the pole positions on sheets II, IV and VIII because, as one  
can see in Ref.~\cite{KMS96}, the analytic continuations of the  
corresponding $S$-matrix elements on these sheets only have the  
forms $$\propto 1/S_{11}^{\rm I},~~~\propto 1/S_{22}^{\rm I}~~ {\rm  
and}~~ \propto 1/S_{33}^{\rm I},$$ respectively. This means that the  
pole positions of resonances only on these sheets are at the same  
points in the complex-energy plane as the resonance zeros on the  
physical sheet and are not shifted due to the coupling of channels.  
Note that the poles on indicated sheets are not always nearest to  
the physical region.  
  
\section{Analysis of the isoscalar-scalar sector }  
  
We analyzed the isoscalar $S$-waves of the processes  
$$\pi\pi\to\pi\pi,K\overline{K},\eta\eta,\eta\eta^\prime.$$  
The experimental data on the $\pi\pi$ scattering  
from 0.575~GeV to 1.89~GeV are taken from  
Ref.~\cite{Hya73} and below 1~GeV from the works~\cite{expd1,expd5,expd6}.  
For $\pi\pi\to K\overline{K}$ the data~\cite{expd2} from threshold  
to about 1.6~GeV are used. The data for $\pi\pi\to\eta\eta$ up to 1.72~GeV  
are taken from Ref.~\cite{expd3} and for $\pi\pi\to\eta\eta^\prime$ from  
threshold to 1.81~GeV from Ref.~\cite{expd4}.  
  
In the model-independent approach we have performed two variants of  
the 3-channel analysis: variant I -- the combined analysis of  
processes $\pi\pi\to\pi\pi,K\overline{K},\eta\eta$; variant II --  
analysis of $\pi\pi\to\pi\pi,K\overline{K},\eta\eta^\prime$.  
The influence of the $\eta\eta^\prime$-channel in the case I and of  
$\eta\eta$ in the case II are taken into account via the background.  
The analysis has been carried out with the new uniformizing  
variable (\ref{w}) ($s_3$ is $4m_\eta^2$ in variant I and  
$(m_\eta+m_{\eta^\prime})^2$ in variant II; in the following the  
quantities related to variant II are primed).  
  
On the $w$-plane the Le Couteur--Newton relations have the  
form~\cite{KMS96}:  
\begin{eqnarray} \label{w:LeCouteur-Newton}  
&&S_{11}=\frac{d^* (-w^*)}{d(w)},~~\qquad  
S_{22}=\frac{d(-w^{-1})}{d(w)},~~\qquad S_{33}=\frac{d(w^{-1})}{d(w)},\\  
&&S_{11}S_{22}-S_{12}^2=\frac{d^*({w^*}^{-1})}{d(w)},\qquad  
S_{11}S_{33}-S_{13}^2=\frac{d^* (-{w^*}^{-1})}{d(w)},\qquad  
S_{22}S_{33}-S_{23}^2=\frac{d(-w)}{d(w)}.\nonumber 
\end{eqnarray}  
In this case the subscripts in the matrix elements  
$S_{\alpha\beta}$ denote  
$\alpha,\beta=$1--$\pi\pi$,~2--$K\overline{K}$,~3--$\eta\eta$ or  
$\eta\eta^{\prime}$.  
  
The $S$-matrix elements in relations (\ref{w:LeCouteur-Newton}) are  
taken as the products~~$S=S_B S_{res}$ where $S_B$ describes the  
background and $S_{res}$ the resonance contributions. The d-function  
for the resonance part is  
\begin{equation} \label{d_res}  
d_{res}(w)=w^{-\frac{M}{2}}\prod_{r=1}^{M}(w+w_{r}^*)  
\end{equation}  
where $M$ is the number of resonance zeros. For the background part $S_B$ the  
$d$-function has the following form:  
\begin{equation}  
d_B=\mbox{exp}[-i(a+\sum_{n=1}^{3}\frac{\sqrt{s-s_n}}{2m_n}  
(\alpha_n+i\beta_n))]  
\end{equation}  
\begin{equation}  
\alpha_n=a_{n1}+a_{n\sigma}\frac{s-s_\sigma}{s_\sigma}\theta(s-s_\sigma)+  
a_{nv}\frac{s-s_v}{s_v}\theta(s-s_v),  
\end{equation}  
\begin{equation}  
\beta_n=b_{n1}+b_{n\sigma}\frac{s-s_\sigma}{s_\sigma}\theta(s-s_\sigma)+  
b_{nv}\frac{s-s_v}{s_v}\theta(s-s_v)  
\end{equation}  
where $s_\sigma$ is the $\sigma\sigma$ threshold, $s_v$ the combined threshold  
of many opened channels in the range  
of $\sim$ 1.5 GeV  
($\eta\eta^{\prime},~\rho\rho,~\omega\omega$). These threshold are  
determined in the analysis.  
  
In variant II, the terms  
$$a_{n\eta}^\prime\frac{s-4m_\eta^2}{4m_\eta^2}  
\theta(s-4m_\eta^2)~~~{\rm and}~~~  
b_{n\eta}^\prime\frac{s-4m_\eta^2}{4m_\eta^2}\theta(s-4m_\eta^2)$$  
were added to $\alpha^\prime_n$ and $\beta^\prime_n$,  
respectively, to account for an influence of the $\eta\eta$--channel.  
  
In the analysis, we included all the  
five resonances discussed below 1.9 GeV in the PDG issue  
\cite{PDG10}. In variant I, for the $\pi\pi$-scattering and  
$\pi\pi\to K\overline{K}$, we considered the data for phase  
shifts and modules of the $S$-matrix elements in the energy regions  
from about 0.4 to 1.89~GeV~ and from the threshold to about 1.6~GeV,  
respectively; for $\pi\pi\to\eta\eta$, the data for the squared  
module of the $S$-matrix element from the threshold to 1.72~GeV. A  
satisfactory description has been obtained.  
Furthermore it was found  
\cite{Kam_up-down} that the data on the $\pi\pi$ scattering below  
1~GeV admit two solutions. Let us call them, as in the indicated  
work \cite{Kam_up-down}, ``up'' and ``down'' because the  
$\pi\pi$-scattering phase shift goes a bit higher in the first case  
than in the second case.  
In variant I, for  
the ``up'' solution, we  
considered the representation of resonances by different  
pole-clusters that are admitted by the data. For the ``down''  
solution, we show the formally best case. In all cases, the  
$f_0(600)$ is represented by the pole cluster of type ({\bf a}), the  
$f_0 (980)$ is represented only by the pole on sheet II and shifted  
pole on sheet III; the resonances $f_0(1370)$ and $f_0(1710)$ can be  
described by the pole clusters of type ({\bf b}) or ({\bf c});  
$f_0(1500)$, of type ({\bf g}).  
  
In Table~\ref{tab:quality}, we demonstrate quality of description  
for each separate process in the frame of the best combined description  
of all three processes for various acceptable variants of  
representation of considered states. We use abbreviations  
$\mbox{``dof''}$ -- number of degrees of freedom and  
$\mbox{``ndp''}$ --  number of data points.  
%
%
\begin{table}[htb!]  
\caption{Variant I: the quality of description of the data  
for the best variants of representation of considered states  
obtained in the analysis. The letters in the second column denote the pole  
clusters describing respectively resonances $f_0(1370)$, $f_0(1500)$  
and $f_0(1710)$.} \label{tab:quality}  
\begin{center}  
\begin{ruledtabular}  
\begin{tabular}{cccccc}  
{Solution} & {} & $\pi\pi$ scattering & $\pi\pi\to K\overline{K}$  
& $\pi\pi\to\eta\eta$ & The total \\  
{} & {} & $\chi^2/\mbox{dof}$ & $\chi^2/\mbox{dof}$ & $\chi^2/\mbox{ndp}$  
& $\chi^2/\mbox{dof}$ \\  
\hline  
{} & bgb & $155.784/(169-35)\approx1.16$ & $148.702/(120-33)\approx1.71$ &  
1.02 & $320.770/(305-42)\approx1.22$ \\  
``up'' & cgb & $150.704/(169-35)\approx1.12$ &  
$154.097/(120-33)\approx1.77$ & 0.91 & $319.447/(305-42)\approx1.21$  
\\  
{} & bgc & $146.684/(169-35)\approx1.09$ &  
$152.274/(120-33)\approx1.75$ & 0.99 & $314.754/(305-42)\approx1.20$  
\\ {} & cgc & $143.269/(169-35)\approx1.07$ &  
$154.469/(120-33)\approx1.78$ & 0.91 & $  
312.365/(305-42)\approx1.19$  
\\  
\hline ``down'' & cgc & $159.425/(169-35)\approx1.19$ &  
$144.475/(120-33)\approx1.66$ & 0.73 & $315.531/(305-42)\approx1.20$  
\end{tabular}  
\end{ruledtabular}  
\end{center}  
\end{table}  
  
It is clear that one should achieve the best description of the  
separate process; however, then the combined description of all  
three processes would be worse.  
  
In Tables~\ref{tab:mass_types} and \ref{tab:mass_down}, there are  
given the masses and total widths of states for the indicated cases,  
calculated from the pole positions on sheets II, IV and VIII for  
resonances of types~({\bf a}),({\bf b})and ({\bf c}), respectively,  
using denominator of the resonance part of amplitude in the  
form~(\ref{T_res}). For the resonance $f_0(1500)$ of type~({\bf g}),  
the poles can be used on all indicated sheets.  
%
%
\begin{table}[htb!]  
\caption{The ``up'' solution: the masses and total widths (in MeV) of  
the $f_0$ resonances, obtained at analyzing for acceptable  
variants of representation of considered states. The letters in the  
upper row denote the pole clusters describing respectively  
resonances $f_0(1370)$, $f_0(1500)$ and $f_0(1710)$.}  
\label{tab:mass_types}  
\begin{center}  
\begin{ruledtabular}  
\begin{tabular}{ccccccccc}  
{} & \multicolumn{2}{c}{bgb} & \multicolumn{2}{c}{cgb} &  
\multicolumn{2}{c}{bgc} & \multicolumn{2}{c}{cgc}\\  
\hline State & $m_{res}$ & $\Gamma_{tot}$ & $m_{res}$ &  
$\Gamma_{tot}$ & $m_{res}$ & $\Gamma_{tot}$ & $m_{res}$  
& $\Gamma_{tot}$ \\  
\hline $f_0(600)$ & 713.7$\pm$5.4 & 627.0$\pm$7.2 & 735.0$\pm$6.1 & 686.0$\pm$7.0  
& 627.0$\pm$7.3 & 665.8$\pm$11.0 & 604.5$\pm$5.7 & 567.0$\pm$5.4 \\  
$f_0 (980)$ & 1007.6$\pm$2.2 & 45.2$\pm$2.8 & 1007.1$\pm$2.6 & 50.6$\pm$2.8  
& 1007.3$\pm$1.9 & 50.8$\pm$2.8 & 1004.7$\pm$ & 54.2$\pm$2.8 \\  
$f_0 (1370)$ & 1404.0$\pm$7.0 & 279.1$\pm$22.0 & 1390.5$\pm$14.3 & 223.4$\pm$42.8 &  
1325.6$\pm$11.1 & 344.6$\pm$24.4 & 1374.5$\pm$16.7 & 322.0$\pm$60.8 \\  
$f_0 (1500)$ & 1532.6$\pm$15.9 & 648.2$\pm$26.6 & 1544.9$\pm$12.2 & 646.2$\pm$26.0  
& 1556.6$\pm$13.5 & 690.4$\pm$28.6 & 1535.4$\pm$12.3 & 671.4$\pm$26.8 \\  
$f_0 (1710)$ & 1750.9$\pm$35.6 & 118.2$\pm$30.2 & 1751.0$\pm$23.8 & 118.0$\pm$50.8  
& 1759.2$\pm$755.7 & 207.0$\pm$420.3 & 1759.2$\pm$716.4 & 201.8$\pm$385.8  
\end{tabular}  
\end{ruledtabular}  
\end{center}  
\end{table}  
%
%
\begin{table}[htb!]  
\caption{The ``down'' solution: the masses and total widths (in MeV)  
of the $f_0$ resonances, obtained at analyzing for the case when the  
resonances $f_0(1370)$, $f_0(1500)$ and $f_0(1710)$ are described by  
the pole clusters of type ({\bf c}), ({\bf g}) and ({\bf c}),  
respectively.} \label{tab:mass_down}  
\begin{center}  
\begin{ruledtabular}  
\begin{tabular}{ccc}  
State  & $m_{res}$ & $\Gamma_{tot}$ \\  
\hline $f_0(600)$ & 769.0$\pm$10.0 & 1036.9$\pm$11.8 \\  
$f_0 (980)$ & 1007.2$\pm$3.1 & 64.6$\pm$3.0 \\  
$f_0 (1370)$ & 1396.4$\pm$24.7 & 355.2$\pm$79.6 \\  
$f_0 (1500)$ & 1534.1$\pm$9.2 & 636.6$\pm$25.8 \\  
$f_0 (1710)$ & 1731.0$\pm$43.6 & 203.4$\pm$34.8  
\end{tabular}  
\end{ruledtabular}  
\end{center}  
\end{table}  
  
As to a quality of description, it is impossible to select any of the  
above-indicated solutions on the basis of analyzing jointly only  
three considered processes. It is required to add in the combined  
analysis also relevant processes of decay. We selected the ``up''  
solution mainly because its parameters of the $f_0 (600)$  
remarkably accord with prediction ($m_\sigma\approx m_\rho$ and  
$\Gamma_{tot}\approx600$ MeV) by Weinberg~\cite{Wei90}, however, for  
now we should consider both solutions. Furthermore, we take a scenario  
in which  the $f_0(1370)$, $f_0(1500)$ and $f_0(1710)$  
are described by the  
pole clusters of type ({\bf c}), ({\bf g}) and ({\bf b}), respectively.  
The point is that the parameters of the $f_0(1500)$ can be calculated  
from the pole positions on sheets II, IV and VIII. Therefore, an  
additional criterion for self-consistency of results is  
a mutual closeness of values of the obtained parameters  
of this important state on the indicated sheets.  
According to this criterion the selected scenario is the most relevant.  
  
In Tables~\ref{tab:clustersI_up} and \ref{tab:clustersI_down}, we  
show the obtained pole clusters for resonances in the complex energy  
plane $\sqrt{s}$, corresponding to the cases when the $f_0(1370)$,  
$f_0(1500)$ and $f_0(1710)$ are described by the pole clusters of the  
type ({\bf c}), ({\bf g}) and ({\bf b}) for the ``up'' and of type  
({\bf c}), ({\bf g}) and ({\bf c}) for ``down'' solutions,  respectively.  
The poles corresponding to the $f_0(1500)$ on sheets IV, VI, and VIII  
are of the 2nd order and those on the sheet V of the 3rd order in our  
approximation.  
\begin{table}[htb!]  
\caption{The pole clusters for the $f_0$-resonances for the ``up''  
solution in variant I. $\!\sqrt{s_r}={\rm E}_r+i\Gamma_r/2\!$~ in  
MeV is given.} \label{tab:clustersI_up}  
\begin{center}  
\begin{ruledtabular}  
{  
\begin{tabular}{ccccccccc}  
{Sheet} & {} & II & III & IV & V & VI & VII & VIII \\  
\hline {\!$f_0 (600)$\!} & {\!${\rm E}_r$\!} & \!650.1$\pm$6.6\! &  
\!703.8$\pm$10.9\! & {} & {} & \!655.9$\pm$27.6\! & \!602.2$\pm$22.0\! & {}  
\\ {} & {\!$\Gamma_r/2$\!} & \!343.0$\pm$3.5\! &  
\!343.0$\pm$3.5\! & {} &{} & \!343.0$\pm$3.5\! & \!343.0$\pm$3.5\! & {}\\  
\hline {\!$f_0(980)$\!} & {\!${\rm  
E}_r$\!} & \!1006.8$\pm$2.6\! & \!980.8$\pm$3.8\! & {} & {} & {} & {} & {}\\  
{} & {\!$\Gamma_r/2$\!} & \!25.3$\pm$1.4 & \!37.2$\pm$2.2\! & {} & {} &  
{} & {} & {}  
\\ \hline {\!$f_0 (1370)$\!} & {\!${\rm E}_r$\!}  
& {} & {} & {} & \!1386.0$\pm$14.2\! & \!1386.0$\pm$14.2\! & \!1386.0$\pm$14.2\! &  
\!1386.0$\pm$14.2\! \\  
{} & {\!$\Gamma_r/2$\!}  
& {} & {} & {} & \!182.9$\pm$34.2\! & \!179.5$\pm$33.6\! & {\!155.7$\pm$18\!} &  
{\!111.7$\pm$21.4\!} \\  
\hline {\!$f_0 (1500)$\!} & {\!${\rm E}_r$\!} & \!1510.7$\pm$12.2\!  
& \!1530.0$\pm$12.7\! & \!1510.7$\pm$12.2\! & \!1510.6$\pm$8.5\! &  
\!1513.$\pm$5.8\! & \!1486.4$\pm$14.6\! & \!1510.7$\pm$12.2\! \\ {} &  
{\!$\Gamma_r/2$\!} & \!323.1$\pm$13.0\! & \!164.7$\pm$11.0\! &  
\!290.5$\pm$46.2\!  
& \!156.9$\pm$9.0\! & \!193.2$\pm$6.4\! & \!134.4$\pm$21.6\! & \!332.9$\pm$73.7\! \\  
\hline {\!$f_0 (1710)$\!} & {\!${\rm E}_r$\!}  
& {} & \!1750.0$\pm$23.8\! & \!1750.0$\pm$23.8\! & \!1750.0$\pm$23.8\!  
& \!1750.0$\pm$23.8\! & {} & {}\\  
{} & {\!$\Gamma_r/2$\!} & {} & \!65.2$\pm$26.6\! & \!59.0$\pm$25.4\! &  
\!63.0$\pm$23.6\! & \!69.2$\pm$24.0\! & {} & {}  
\end{tabular}  
}  
\end{ruledtabular}  
\end{center}  
\end{table}  
\begin{table}[htb!]  
\caption{The pole clusters for the $f_0$-resonances for the ``down''  
solution in variant I. $\!\sqrt{s_r}={\rm E}_r+i\Gamma_r/2\!$~ in  
MeV is given.} \label{tab:clustersI_down}  
\begin{center}  
\begin{ruledtabular}  
{  
\begin{tabular}{ccccccccc}  
{Sheet} & {} & II & III & IV & V & VI & VII & VIII \\  
\hline {\!$f_0 (600)$\!} & {\!${\rm E}_r$\!} & \!567.9$\pm$12.4\! &  
\!642.0$\pm$17.7\! & {} & {} & \!647.7$\pm$29.1\! & \!573.6$\pm$25.5\! &  
{}  
\\ {} & {\!$\Gamma_r/2$\!} & \!518.5$\pm$5.9\! &  
\!518.5$\pm$5.9\! & {} &{} & \!518.5$\pm$5.9\! & \!518.5$\pm$5.9\! & {}\\  
\hline {\!$f_0(980)$\!} & {\!${\rm  
E}_r$\!} & \!1006.7$\pm$3.1\! & \!970.1$\pm$5.8\! & {} & {} & {} & {} & {}\\  
{} & {\!$\Gamma_r/2$\!} & \!32.3$\pm$1.5 & \!55.4$\pm$2.6\! & {} & {} &  
{} & {} & {}  
\\ \hline {\!$f_0 (1370)$\!} & {\!${\rm E}_r$\!}  
& {} & {} & {} & \!1385.1$\pm$24.4\! & \!1385.1$\pm$24.4\! &  
\!1385.1$\pm$24.4\! & \!1385.1$\pm$24.4\! \\  
{} & {\!$\Gamma_r/2$\!} & {} & {} & {} & \!287.0$\pm$73.7\! &  
\!267.4$\pm$83.1\! & {\!158.0$\pm$41.8\!} &  
{\!177.6$\pm$39.8\!} \\  
\hline {\!$f_0 (1500)$\!} & {\!${\rm E}_r$\!} & \!1500.7$\pm$4.9\! &  
\!1495.0$\pm$9.0\! & \!1500.7$\pm$4.9\! & \!1496.7$\pm$7.2\! &  
\!1510.2$\pm$4.6\! & \!1501.2$\pm$9.8\! & \!1500.7$\pm$4.9\! \\ {} &  
{\!$\Gamma_r/2$\!} & \!318.3$\pm$12.9\! & \!133.6$\pm$10.6\! &  
\!231.9$\pm$17.6\!  
& \!141.$\pm$6.3\! & \!185.2$\pm$4.0\! & \!99.$\pm$18.0\! & \!345.9$\pm$14.5\! \\  
\hline {\!$f_0 (1710)$\!} & {\!${\rm E}_r$\!} & {} & {} & {} & \!1728.0$\pm$43.7\!  
& \!1728.0$\pm$43.7\! & \!1728.0$\pm$43.7\! & \!1728.0$\pm$43.7\! \\  
{} & {\!$\Gamma_r/2$\!} & {} & {} & {} & \!139.9$\pm$69.0\! & \!138.7$\pm$8.9\!  
& \!100.5$\pm$48.4\! & \!101.7$\pm$17.4\!  
\end{tabular}  
}  
\end{ruledtabular}  
\end{center}  
\end{table}  
  
The background parameters are: ``up'' solution -- $a=0.4704\pm0.0364$,  
$a_{11}=-0.2376\pm0.0132$, $a_{1\sigma}=0.186\pm0.0335$, $a_{1v}=-0.0788\pm0.0535$,  
$b_{11}=b_{1\sigma}=0$, $b_{1v}=0.0305\pm0.0112$, $a_{21}=-1.7768\pm0.0461$,  
$a_{2\sigma}=0.5204\pm0.0254$, $a_{2v}=-9.22\pm0.649$, $b_{21}=0.0132\pm0.0131$,  
$b_{2\sigma}=0$, $b_{2v}=7.385\pm1.354$, $b_{31}=0.5494\pm0.0458$,  
$b_{3\sigma}=0.8995\pm0.0997$, $b_{3v}=0$; $s_\sigma=1.638~{\rm GeV}^2$,  
$s_v=2.085~{\rm GeV}^2$; ``down'' solution -- $a=0.2431\pm0.0322$,  
$a_{11}=-0.0553\pm0.0113$, $a_{1\sigma}=0.0914\pm0.0103$,  
$a_{1v}=-0.0478\pm0.0098$,  
$b_{11}=b_{1\sigma}=0$, $b_{1v}=0.0469\pm0.0104$, $a_{21}=-1.6811\pm0.0426$,  
$a_{2\sigma}=-0.247\pm0.1987$, $a_{2v}=-7.2\pm0.5858$, $b_{21}=0.0329\pm0.0131$,  
$b_{2\sigma}=0$, $b_{2v}=7.765\pm1.4301$, $b_{31}=0.6135\pm0.0495$,  
$b_{3\sigma}=0.6617\pm0.1099$, $b_{2v}=0$.  
  
The obtained zero positions of the resonances on the $w$-plane are: \\  
``up'' solution --  
\begin{eqnarray}  
{\rm for}~f_0(600): &w_1=3.8538993 + 4.1546473i,  
&w_2=-0.1320698 + 0.1400331i,\nonumber\\  
&w_3=-3.8192504 - 4.1137414i, &w_4=0.1109493 - 0.1194408i,\nonumber\\  
{\rm for}~f_0(980): &w_5=0.6671266 + 1.1471002i,  
&w_6=-0.2290429 + 0.5613030i,\nonumber\\  
{\rm for}~f_0(1370): &w_7=-3.0340738 - 0.4551088i,  
&w_8=3.0385254 - 0.4624306i,\nonumber\\  
&w_9=-0.3346148 - 0.0335710i, &w_{10}=0.3351386 -0.0326940i,\nonumber\\  
{\rm for}~f_0(1500): &w_{11}=3.4148479 + 0.5291918i,  
&w_{12}=0.2899891 +0.0420199i,\nonumber\\  
&w_{13}=-0.2847651 - 0.0449140i,  
&w_{14}=w_{15}=-0.3016876 + 0.0264597i,\nonumber\\  
w_{16}=&w_{17}=w_{18}=-3.2842692 - 0.3472515i,&{}\nonumber\\  
&w_{19}=w_{20}=3.2517666 - 0.2894345i,  
&w_{21}=w_{22}=0.3110365 - 0.0259720i,\nonumber\\  
{\rm for}~f_0(1710): &w_{23}=0.2818700 + 0.0051870i,  
&w_{24}=-0.2817805 + 0.0057246i.\nonumber\\  
&w_{25}=-3.5479990 - 0.0764492i, &w_{26}=3.5470806 - 0.0696555i;\nonumber  
\end{eqnarray}  
``down'' solution --  
\begin{eqnarray}  
{\rm for}~f_0(600): &w_1=5.2482964 + 3.3618022i,  
&w_2=-0.1459568 + 0.0959815i,\nonumber\\  
&w_3=-4.7510997 - 3.1267321i, &w_4=0.1358737 - 0.0873064i,\nonumber\\  
{\rm for}~f_0(980): &w_5=0.7366709 + 1.2022465i,  
&w_6=-0.2172355 + 0.4780125i,\nonumber\\  
{\rm for}~f_0(1370): &w_7=-3.1626888 - 0.6250175i,  
&w_8=3.1950376 - 0.6565649i,\nonumber\\  
&w_9=-0.3228711 - 0.0481947i, &w_{10}=0.3267097 - 0.0444134i,\nonumber\\  
{\rm for}~f_0(1500): &w_{11}=3.3973448 + 0.5354069i,  
&w_{12}=0.2981317 + 0.0378036i,\nonumber\\  
&w_{13}=-0.2837024 - 0.0468876i,  
&w_{14}=w_{15}=-0.3096900 + 0.0250295i,\nonumber\\  
w_{16}=&w_{17}=w_{18}=-3.2730730 - 0.3367541i,&{}\nonumber\\  
&w_{19}=w_{20}=3.2161986 - 0.2715142i,  
&w_{21}=w_{22}=0.3114429 - 0.0186598i,\nonumber\\  
{\rm for}~f_0(1710): &w_{23}=-3.5403847 - 0.1579625i,  
&w_{24}=3.5407727 - 0.1592757i.\nonumber\\  
&w_{25}=-0.2829782 - 0.0093685i, &w_{26}=0.2830083 - 0.0092620i.\nonumber  
\end{eqnarray}  
  
For variant II we got the following description: for $\pi\pi$  
scattering $\chi^2/\mbox{dof}=148.786/(169-30)\approx 1.07$;  
for~$\pi\pi\to K\overline{K}$  
$\chi^2/\mbox{dof}=155.006/(120-29)\approx1.70$;  
for~$\pi\pi\to\eta\eta^\prime$  $\chi^2/\mbox{ndp}\approx0.3$. The  
total $\chi^2/\mbox{dof}$ is $306.187/(297-37)\approx 1.18$. In this  
case the $f_0 (600)$ is described by the cluster of type ({\bf  
a}$^\prime$); $f_0 (1370)$, type ({\bf b}$^\prime$); $f_0 (1500)$,  
type ({\bf d}$^\prime$); $f_0(1710)$, type ({\bf c}$^\prime$). In  
Table~\ref{tab:clustersII} we indicate the obtained pole clusters  
for resonances on the eight sheets in the complex energy plane  
$\sqrt{s}$. The poles on sheets IV and V, corresponding to the  
$f_0(1500)$, are of the 2nd order (this is an approximation).  
%
%
\begin{table}[htb]  
\caption{The pole clusters for the $f_0$-resonances in variant II.  
$\!\sqrt{s_r^{\prime}}={\rm E}_r^{\prime}+i\Gamma_r^{\prime}/2\!$~  
in MeV is given.} \label{tab:clustersII}  
\begin{center}  
\begin{ruledtabular}  
{  
\begin{tabular}{ccccccccc}  
{Sheet} & {} & II & III & IV & V & VI & VII & VIII \\ \hline {\!$f_0  
(600)$\!} & {\!${\rm E}_r^{\prime}$\!} & \!558.7$\pm$13.3\! &  
\!564.3$\pm$13.7\! & {} & {} & \!541.3$\pm$55.0\! & \!535.7$\pm$26.0\! & {}  
\\ {} & {\!$\Gamma_r^{\prime}/2$\!} & \!529$\pm$17.4\! & \!529$\pm$17.4\! & {}  
&{} & \!529$\pm$17.4\! &  
\!529$\pm$17.4\! & {}\\  
\hline {\!$f_0(980)$\!} & {\!${\rm E}_r^{\prime}$\!} &  
\!1009.0$\pm$3.1\! & \!986.1$\pm$5.5\! & {} & {} & {} & {} & {}\\ {} &  
{\!$\Gamma_r^{\prime}/2$\!} &  
\!31.8$\pm$1.8\! & \!57.4$\pm$2.9\! & {} & {} & {} & {} & {} \\  
\hline {\!$f_0 (1370)$\!} & {\!${\rm E}_r^{\prime}$\!} & {} &  
\!1411.6$\pm$8.1\! & \!1411.6$\pm$8.1\! & \!1428.4$\pm$11.0\! &  
\!1428.4$\pm$11.0\! & {} & {} \\  
{} & {\!$\Gamma_r^{\prime}/2$\!} & {} & \!215.6$\pm$21.2\! &  
\!235$\pm$22.6\! & \!235$\pm$22.6\! &  
\!215.6$\pm$21.2\! & {} & {} \\  
\hline {\!$f_0 (1500)$\!} & {\!${\rm E}_r^{\prime}$\!} &  
\!1496.9$\pm$4.7\! & \!1503.0$\pm$3.9\! & \!1496.9$\pm$4.7\! &  
\!1496.9$\pm$4.7\! & \!1494.6$\pm$2.9\! & \!1496.9$\pm$4.7\! & {} \\ {}  
& {\!$\Gamma_r^{\prime}/2$\!} & \!198.5$\pm$7.8\! & \!236.0$\pm$5.7\! &  
\!193.1$\pm$8.6\! & \!198.5$\pm$7.8\! & \!193.7$\pm$4.7\! &  
\!193.1$\pm$8.6\! & {} \\  
\hline {\!$f_0 (1710)$\!} & {\!${\rm E}_r^{\prime}$\!} & {} & {} &  
{} & \!1743.0$\pm$17.8\! & \!1743.0$\pm$17.8\! & \!1743.0$\pm$17.8\! &  
\!1743.0$\pm$17.8\! \\  
 & {\!$\Gamma_r^{\prime}/2$\!}  
& {} & {} & {} & \!144.1$\pm$40.3\! & \!111.5$\pm$31.9\! & \!82.1$\pm$36.6\!  
& \!114.7$\pm$38.6\!  
\end{tabular}  
}  
\end{ruledtabular}  
\end{center}  
\end{table}  
  
The background parameters are: $a^\prime=0.2315\pm0.0085$, $a_{11}^\prime=0$,  
$a_{1\eta}^\prime=-0.0616\pm0.0321$, $a_{1\sigma}^\prime=0.0298\pm0.0876$,  
$a_{1v}^\prime=0.0622\pm0.0703$,  
$b_{11}^\prime=b_{1\eta}^\prime=b_{1\sigma}^\prime=0$,  
$b_{1v}^\prime=0.0449\pm0.0105$, $a_{21}^\prime=-3.1359\pm0.0628$,  
$a_{2\eta}^\prime=0$, $a_{2\sigma}^\prime=0.4866\pm0.2778$,  
$a_{2v}^\prime=-4.532\pm0.7199$, $b_{21}^\prime=0$,  
$b_{2\eta}^\prime=-0.7478\pm0.0607$, $b_{2\sigma}^\prime=2.5545\pm0.2067$,  
$b_{2v}^\prime=1.948\pm1.785$, $b_{31}^\prime=0.4489\pm0.0606$,  
$s_\sigma=1.638~{\rm GeV}^2$, $s_v=2.126~{\rm GeV}^2$.  
  
The obtained zero positions of the resonances on the $w^\prime$--plane are:  
\begin{eqnarray}  
{\rm for}~f_0(600): &w_1^{\prime}=2.8974465 + 2.0075214i,  
&w_2^{\prime}=-0.2343026 + 0.1631268i,\nonumber\\  
&w_3^{\prime}=-2.9703646 - 2.0251533i,  
&w_4^{\prime}=0.2287780 - 0.1551171i,\nonumber\\  
{\rm for}~f_0(980): &w_5^{\prime}=0.3793973 + 1.1246624i,  
&w_6^{\prime}=-0.2267376 + 0.6892463i,\nonumber\\  
{\rm for}~f_0(1370): &w_7^{\prime}=0.6466646 + 0.2560839i,  
&w_8^{\prime}=-0.6616945 + 0.2603249i,\nonumber\\  
&w_9^{\prime}=-1.3218615 - 0.4907520i,  
&w_{10}^{\prime}=1.3490514 - 0.5056393i,\nonumber\\  
{\rm for}~f_0(1500): &w_{11}^{\prime}=1.3583321 + 0.3852687i,  
&w_{12}^{\prime}=0.6855763 + 0.1928257i,\nonumber\\  
&w_{13}^{\prime}=1.3583321 - 0.3852687i,  
&w_{14}^{\prime}=0.6855763 - 0.1928257i,\nonumber\\  
w_{15}^{\prime}=&w_{16}^{\prime}=-0.6543233 + 0.1907843i,  
w_{17}^{\prime}=&w_{18}^{\prime}=-1.3502545 - 0.3835908i,\nonumber\\  
{\rm for}~f_0(1710): &w_{19}^{\prime}=-1.5488743 - 0.1115869i,  
&w_{20}^{\prime}=1.5602461 - 0.1413653i,\nonumber\\  
&w_{21}^{\prime}=-0.6417003 - 0.0474415i,  
&w_{22}^{\prime}=0.6471581 - 0.0350139i.\nonumber  
\end{eqnarray}  
  
Masses and total widths of the states, calculated from the pole  
positions on sheets II , IV and VIII for resonances of types~({\bf  
a}),({\bf b})and ({\bf c}), respectively, and on sheets II or IV for  
resonance of type~({\bf d}), are presented in Table~\ref{tab:massII}.  
%
%
\begin{table}[htb!]  
\caption{Variant II: the masses and total widths of the $f_0$  
resonances.} \label{tab:massII}  
\begin{center}  
\begin{ruledtabular}  
{  
\begin{tabular}{ccccc}  
{State} & $m_{res}^{\prime}$ [MeV] & $\Gamma_{tot}^{\prime}$ [MeV] \\  
\hline $f_0 (600)$ & 769.4$\pm15.4$ & 1058.0$\pm34.8$ \\  
$f_0 (980)$ & 1009.5$\pm3.1$ & 63.6$\pm3.6$ \\  
$f_0 (1370)$ & 1431.0$\pm8.8$ & 469.9$\pm45.2$ \\  
$f_0 (1500)$ & 1510.1$\pm4.8$ & 397.1$\pm15.6$ \\  
$f_0 (1710)$ & 1746.8$\pm17.9$ & 229.4$\pm77.2$  
\end{tabular}  
}  
\end{ruledtabular}  
\end{center}  
\end{table}  
  
\begin{figure}[htb]  
\begin{center}  
\includegraphics[width=0.5\textwidth,angle=-90]{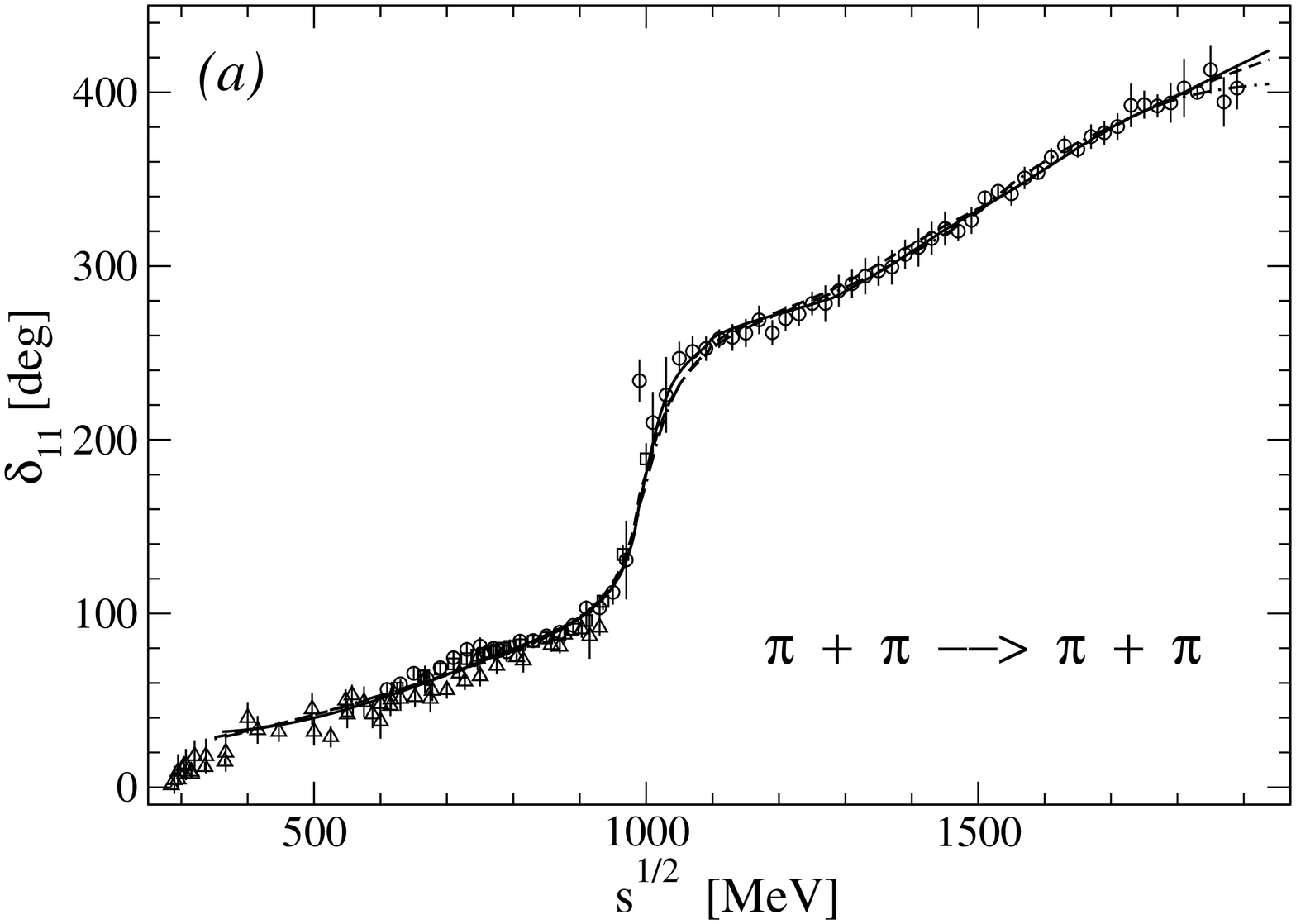}  
\includegraphics[width=0.5\textwidth,angle=-90]{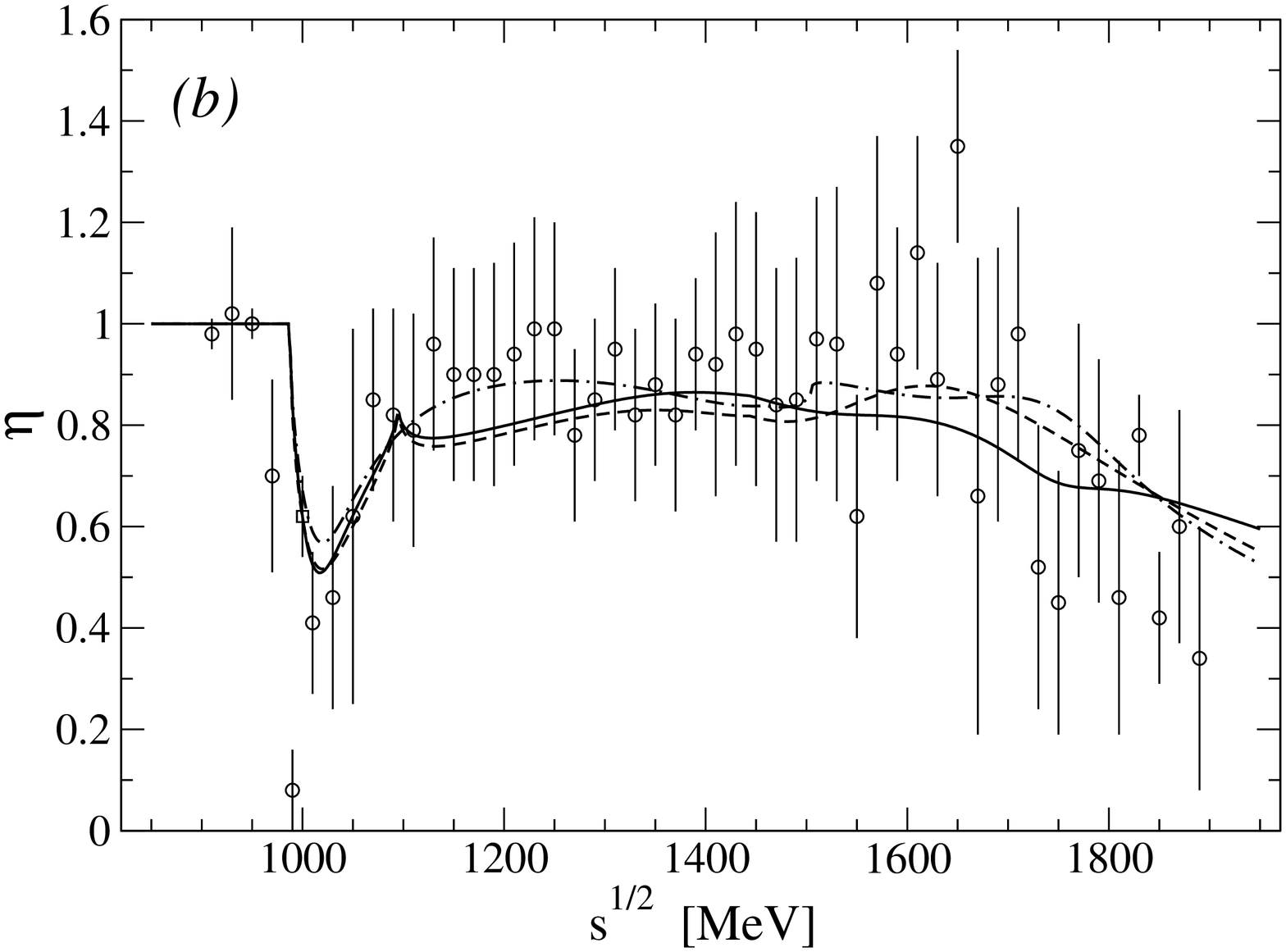}  
\caption{The phase shift and modulus of the $S$-matrix element in the  
$S$-wave $\pi\pi$-scattering. The solid and short-dashed curves  
correspond to variant I, the ``up'' and ``down'' solutions,  
respectively; dash-dotted to variant II. The data are from  
Refs.~\cite{Hya73,expd1,expd5,expd6}.} \label{fig:S11}  
\end{center}  
\end{figure}

\begin{figure}[htb]  
\begin{center}  
\includegraphics[width=0.5\textwidth,angle=-90]{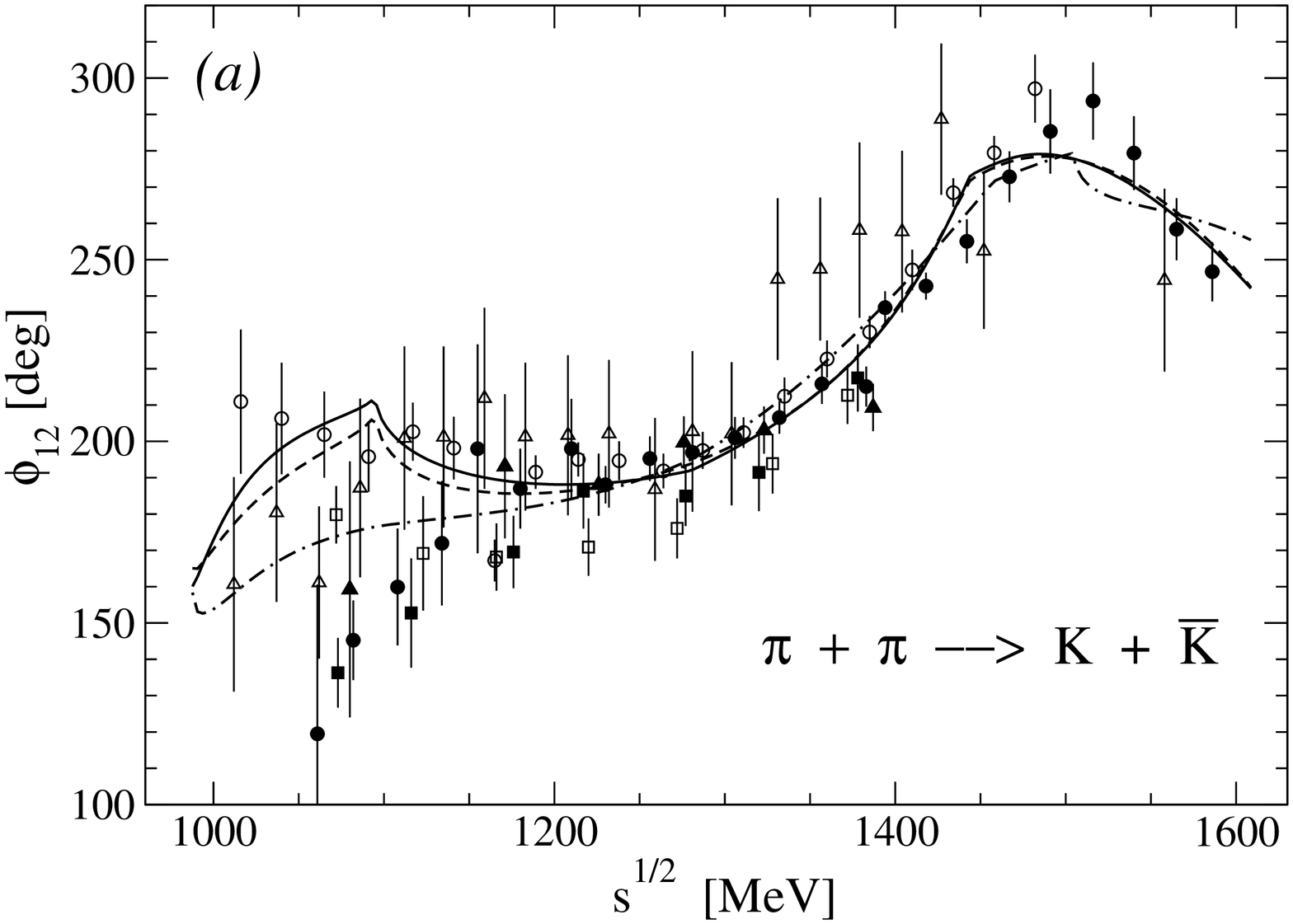}  
\includegraphics[width=0.5\textwidth,angle=-90]{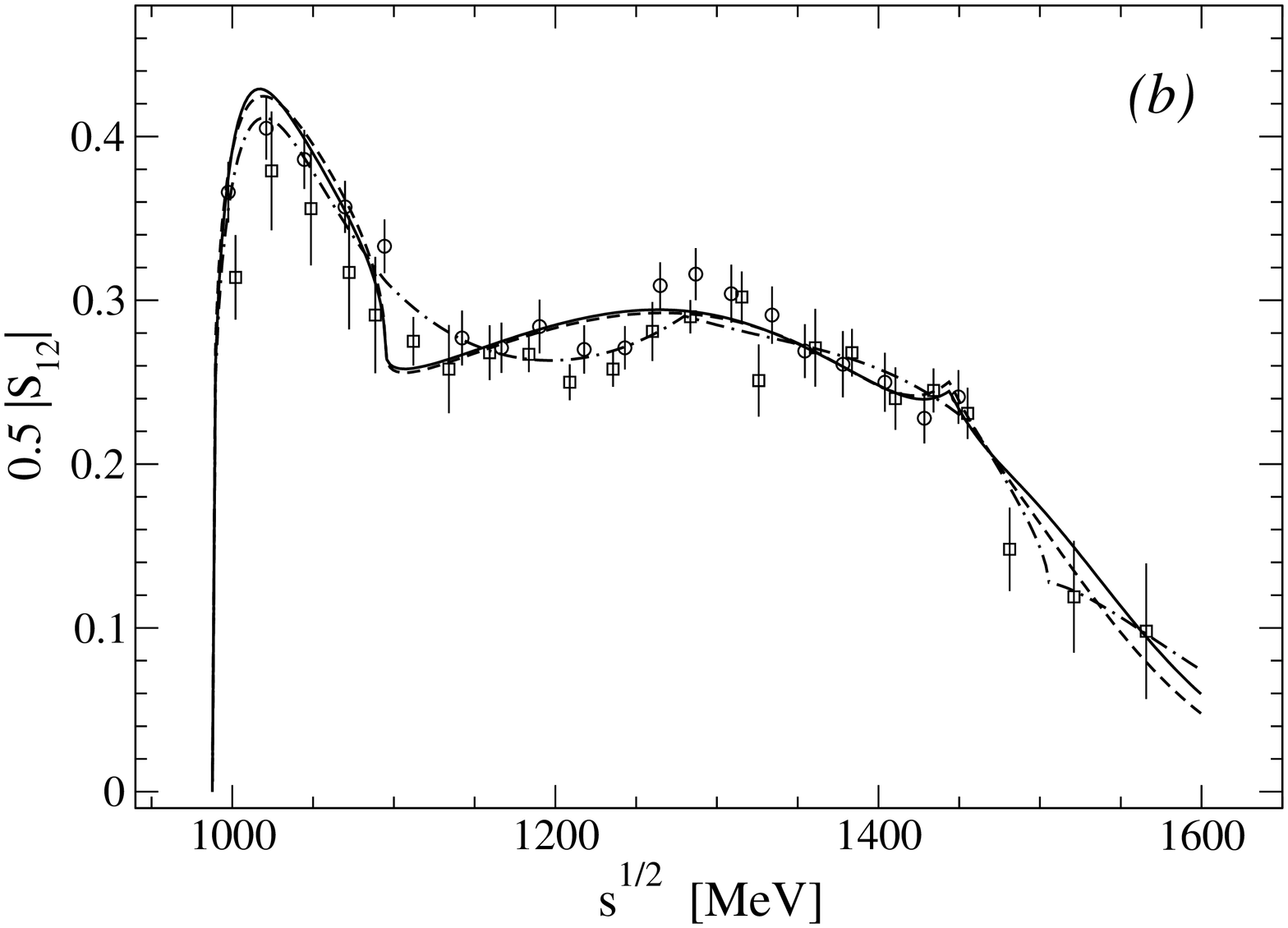}  
\caption{The phase shift and modulus of the $S$-matrix element in  
$S$-wave of $\pi\pi\to K\overline{K}$. The solid and short-dashed  
curves correspond to variant I, the ``up'' and ``down'' solutions,  
respectively; dash-dotted to variant II. The data are from  
Ref.~\cite{expd2}.} \label{fig:S12}  
\end{center}  
\end{figure}

\begin{figure}[htb]  
\begin{center}  
\includegraphics[width=0.5\textwidth,angle=-90]{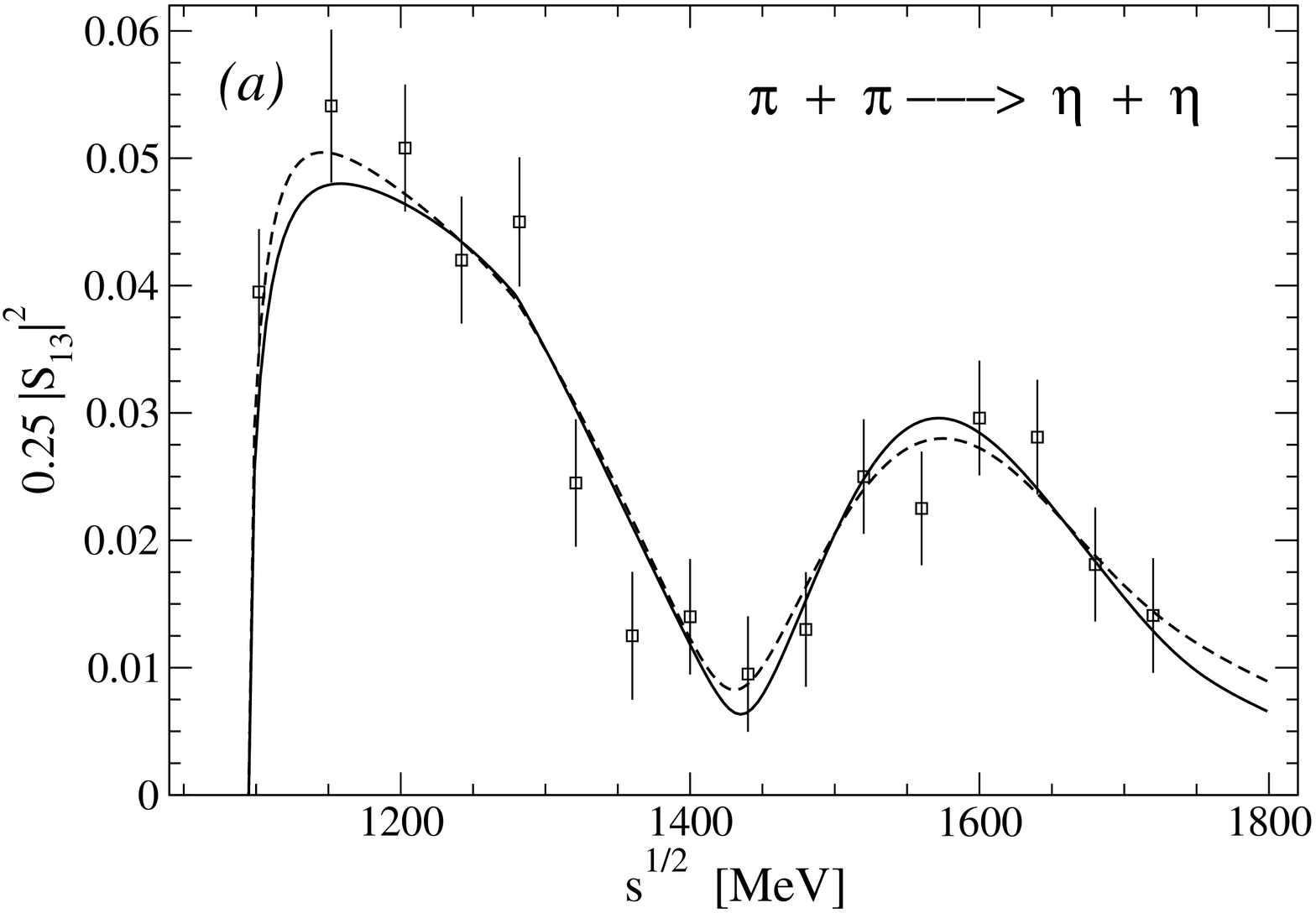}  
\includegraphics[width=0.5\textwidth,angle=-90]{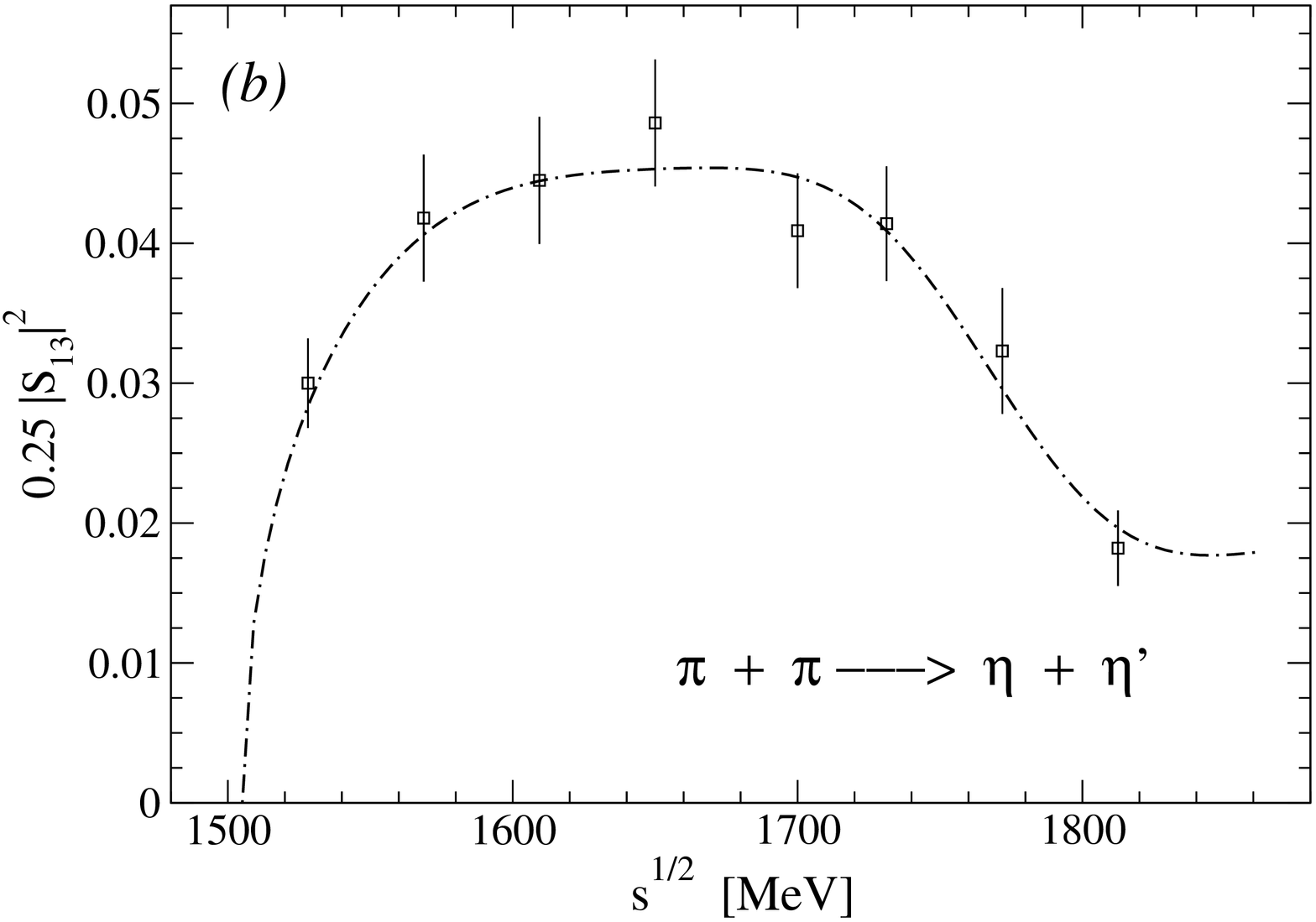}  
\caption{The squared modules of the $\pi\pi\to\eta\eta$ (upper  
figure) and $\pi\pi\to\eta\eta^\prime$ (lower figure) $S$-wave  
matrix elements. The data are  
from Ref.~\cite{expd3} (upper figure)  
and  
from Ref.~\cite{expd4} (lower figure).} \label{fig:S13}  
\end{center}  
\end{figure}

In Figures~\ref{fig:S11}--\ref{fig:S13}, we show results of fitting  
to the experimental data in both variants.  
%
%
%
%
%
%
  
Note, that  
the ``up'' solution is not revealed in variant II when in the analysis  
the data both below and above the $K\overline{K}$-threshold are used.  
This suggests that, the  
$\eta\eta$-threshold branch-point must be taken  
into account explicitly. It is not sufficient  
to consider influence of the $\eta\eta$ channel only via the background.  
  
To learn more on existence of the $f_0 (1370)$  
(see the discussion in Introduction),  
we considered a possibility of description in the above-selected  
cases without this state, i.e. when the resonances $f_0(600)$,  
$f_0(1500)$, and $f_0(1710)$ are represented respectively by the pole  
clusters of type ({\bf a}), ({\bf g}) and ({\bf b}) in variant I (the  
``up'' solution) and  
of type ({\bf a}$^\prime$), ({\bf d}$^\prime$) and ({\bf c}$^\prime$) in  
variant II. The $f_0 (980)$ is represented by the poles on sheets II  
and III in both variants. In Table~\ref{tab:quality_without} we  
show a quality of description of each separate process for these cases in  
the frame of the best combined description of all three processes.  
%
%
\begin{table}[htb!]  
\caption{The quality of description of the data without the  
$f_0(1370)$.} \label{tab:quality_without}  
\begin{center}  
\begin{ruledtabular}  
\begin{tabular}{ccccc}  
Variant & $\pi\pi$ scattering & $\pi\pi\to K\overline{K}$  
& $\pi\pi\to\eta\eta,\eta\eta^\prime$ & The total \\  
{} & $\chi^2/\mbox{dof}$ & $\chi^2/\mbox{dof}$ &  
$\chi^2/\mbox{ndp}$ & $\chi^2/\mbox{dof}$ \\  
\hline I & $151.395/(169-31)\approx1.10$ &  
$155.056/(120-29)\approx1.70$ & 1.04 & $323.14/(305-38)\approx1.21$  
\\  
\hline II & $150.145/(169-26)\approx1.05$ &  
$160.056/(120-26)\approx1.70$ & 0.38 & $313.225/(305-34)\approx1.19$  
\end{tabular}  
\end{ruledtabular}  
\end{center}  
\end{table}  
  
When calculating $\chi^2$ in all cases, the following experimental  
points have been omitted as obviously strongly falling out from the  
energy dependence: from the $\pi\pi$ scattering data the points at  
990~MeV for the phase shift $\delta_{11}$ and for $\eta_{11}=|S_{11}|$,  
and for $\eta_{11}$ the point at 1650~MeV as strongly violating the  
unitarity condition. From the $\pi\pi\to K\overline{K}$ data there were  
omitted the points at 1002, 1208.9 and 1235.7~MeV for the  
$\eta_{12}=|S_{12}|$ and the points at 1073, 1082 and 1387~MeV as giving  
the anomalously big contribution to $\chi^2$.  
  
One can see that the description of the $\pi\pi$ scattering without the  
$f_0(1370)$ is a bit improved whereas the one of the  
$\pi\pi\to K\overline{K}$ process is made slightly worse, especially as  
to the phase shift. Generally, an existence of the $f_0(1370)$ is for  
now a standard point of view.  
One ought to take into account also arguments to its favor in Ref.  
\cite{Bugg1370} (see Introduction). In any case, the existence of  
the $f_0(1370)$ does not contradict the considered data.  
  
Let us make some more remarks. First, the fact that in variant II we  
obtain a better description than in variant I points to the importance  
of taking into account the $\eta\eta^\prime$ threshold explicitly.  
However, in variant II we encounter elements of some pseudo-background:  
these are the negative values of the $b$ coefficients related to an  
inelastic part of the background. The increasing inelastic part of the  
background implies a necessity to consider explicitly some physical  
phenomenon, e.g., additional resonances or representation of resonances  
by other pole-clusters or the consideration in the uniformizing  
variable of other channel thresholds. The latter situation is the case  
here: the negative sign of the quantity $b_{2\eta}^\prime=-0.7478$  
implies the necessity of an explicit consideration of the  
$\eta\eta$-threshold branch-point. Therefore, as to the resonances  
below 1500~MeV the more adequate description is variant I whereas for  
the ones above 1500~MeV variant II.  
  
It turns out that the state $f_0(980)$ lies slightly above the  
$K\overline{K}$ threshold. It is described by the pole on sheet II  
and by the shifted pole on sheet III under the $\eta\eta$ threshold  
without the corresponding poles on sheets VI and VII, as it was  
expected for standard clusters. This may suggest that the $f_0(980)$  
is not the $q{\bar q}$ state and can be interpreted, e.g., as a  
$\eta\eta$ bound state in accordance with the test discussed in Sec.II:  
the necessary condition for this is fulfilled. See, however, the  
further discussion of this matter in the last section.  
  
As to a representation of the $f_0(600)$ and $f_0(980)$ states, both  
variants completely agree. The $f_0(1370)$ is described by the clusters of  
type ({\bf b}) or ({\bf c}) in various scenarios of variant I and of type  
({\bf b}$^\prime$) in variant II; this is reasonable taking into account  
the quark contents of the $K\overline{K}$ and $\eta\eta$ systems and the  
nearness of corresponding thresholds. From this we, therefore, deduce 
that a $s{\bar s}$ component of the $f_0(1370)$ is dominant. 
This interpretation quite  
explains why one did not find evidence for the existence of the  
$f_0(1370)$ \cite{Ochs10} considering only the $\pi\pi$ scattering.  
  
The $f_0(1500)$ is described by the cluster of type ({\bf g}) in variant I  
and of type ({\bf d}$^\prime$) in variant II. The former indicates  
approximately equal coupling constants of this state to the $\pi\pi$,  
$K\overline{K}$ and $\eta\eta$ systems, which apparently could point up to its  
glueball nature. The latter tells on the approximately equal coupling of this  
state with the $\pi\pi$ and $K\overline{K}$ channels whereas the coupling with  
the $\eta\eta^{\prime}$ channel is suppressed; these facts also point up to  
its glueball nature \cite{Ams96}. Therefore, we deduce a dominant glueball  
component of the $f_0(1500)$.  
  
Finally, the $f_0(1710)$ is described by the clusters of type ({\bf b}) or  
({\bf c}) in various scenarios of variant I and of type ({\bf c}$^\prime$) in  
variant II. Taking also into account the quark contents of the  
$\eta\eta^{\prime}$ system this could point to the dominant $s{\bar s}$  
component of this state.  
  
All these conclusions agree quite well with the previous  
model-independent 2- and 3-channel  
analyses~\cite{Sur-Kam_07,SBKN-ijmp09,SBKN-prd10,SKN-epja,%
SKN-AIP04,PRD-01,SKN-czjp06,KMS96}  
where other uniformizing variables were used.  
  
\section{Discussion and conclusions}  
The combined analysis of data on the  
$\pi\pi\to\pi\pi,K\overline{K},\eta\eta,\eta\eta^\prime$ processes in the  
channel with $I^GJ^{PC}=0^+0^{++}$ is carried out in the framework of the  
model-independent approach that is based on analyticity and unitarity and  
uses an uniformization procedure. A new uniformizing variable was used in  
which, in additional to the right-hand branch-points related with the  
thresholds of the analyzed channels, there is taken into account the  
left-hand branch-point at $s=0$ related to the $\pi\pi$ scattering in the  
crossed channels.  
  
In the analysis of the processes  
$\pi\pi\to\pi\pi,K\overline{K},\eta\eta$ it is shown that the data admit  
two possibilities for parameters of the $f_0(600)$ with mass, relatively near  
to the $\rho$-meson mass, and with total width about 640 and 1000 MeV. These  
two possibilities are related to two found solutions, admissible by the data  
below 1 GeV for the phase shift of the $\pi\pi$-scattering amplitude: ``up'' and  
``down''. As to the combined description of the considered processes  
it is impossible to prefer any of these solutions. However, the ``up'' solution  
remarkably accord with prediction by Weinberg~\cite{Wei90}  
with respect to the mass and the width, the ``down'' one to the mass.  
These values of mass and width, calculated with  
help of formula (\ref{T_res})  
from the pole position on sheet II, correspond to most of the Breit--Wigner  
values of Refs. \cite{Tornqvist96} (analysis of several processes with  
pseudoscalar mesons) and \cite{Alde97} (GAMS Collaboration, analysis of the  
reaction $pp\to pp\pi^0\pi^0$).  
  
Furthermore, we have considered all relevant possibilities of  
representation of resonances by pole clusters (the 3-channel resonances  
are represented by seven types of the pole clusters).  
It is shown that for the ``up'' solution there are four scenarios of  
representation of resonances $f_0(1370)$, $f_0(1500)$ (as the  
superposition of two states, broad and narrow) and $f_0(1710)$ ($f_0(600)$  
and $f_0(980)$ are given by the pole clusters of the same types in all cases)  
giving about the similar description of the above processes and, however,  
the quite different parameters of some resonances. For the $f_0(600)$,  
$f_0(1370)$ and $f_0(1710)$ a spread of values is obtained for the masses  
and widths 605-735 and 567-686~MeV, 1326-1404 and 223-345~MeV, and  
1751-1759 and 118-207~MeV, respectively. On the other hand, the results  
for the $f_0(980)$ and $f_0(1500)$ are more stable and confirm conclusions  
of our previous analyses~\cite{Sur-Kam_07,SBKN-ijmp09,SBKN-prd10,%
SKN-epja,SKN-AIP04,PRD-01,SKN-czjp06}.  
  
Note a quite stable result for the mass and  
width with rather small errors for  
the $f_0(980)$: $m_{res}\approx1005-1008$~MeV, $\Gamma_{tot}\approx45-54$~MeV.  
Arrangement of the poles and zeroes on the Riemann surface, which describe  
this state, may suggest that the $f_0(980)$ is not the $q{\bar q}$ state and  
can be interpreted, e.g., as a $\eta\eta$ bound state; in any case the  
necessary condition for this is fulfilled.  
However, following the listings PDG \cite{PDG10}, the mass of this state  
is obtained above the $K\overline{K}$ threshold in analyses of $\pi\pi$  
scattering, of multi-channel $\pi\pi$ scattering  
($\pi\pi\to\pi\pi,K\overline{K},\eta\eta,\eta\eta^\prime$) and of processes  
${\bar p}p(n)\to M_1M_2M_3$, whereas below the $K\overline{K}$ threshold in  
analyses of the decays of $D^+$--, $B^+$--, $J/\psi$--, and $Z$--bosons, of  
processes $e^+e^-\to M_1M_2\gamma, \phi M_1M_2\gamma,e^+e^-M_1M_2,M_1M_2X$,  
and of $pp\to ppM_1M_2$. Since the mass value below the $K\overline{K}$  
threshold is important for a dynamical interpretation of the $f_0(980)$ as  
a $K\overline{K}$ molecule~\cite{Isgur,Jansen,BGL} it seems that the  
nature of this state is more complicated than a simple $\eta\eta$ bound  
state or $K\overline{K}$ molecule. From the point of view of the quark  
structure these two possibilities are the 4-quark states.  
It seems this is consistent somehow with arguments in favor of the  
4-quark nature of $f_0(980)$ in work of \cite{Achasov00}.  
  
In view of prolonging discussions of a question, whether the $f_0 (1370)$  
exists or not (see the discussion of this matter in Introduction),  
we considered a description of the  
multi-channel $\pi\pi$ scattering without this state.  
We concluded that an existence  
of the $f_0(1370)$ does not contradict the  
considered data. The description of the $\pi\pi$ scattering is a bit improved  
whereas the one of the $\pi\pi\to K\overline{K}$ process is made  
worse, especially as to the phase shift.  
  
The ${f_0}(1370)$ (if it exists) and $f_0 (1710)$ have  
a dominant $s{\bar s}$ component.  
Conclusion about the ${f_0}(1370)$ agrees quite well with the conclusion  
drawn by the Crystal Barrel Collaboration~\cite{15} where the ${f_0}(1370)$  
is identified as $\eta\eta$ resonance in the $\pi^0\eta\eta$ final state of  
the ${\bar p}p$ annihilation at rest. Interpretation of the $f_0(1370)$  
as dominated by the $s{\bar s}$ component explains also quite well  
why one did not find this state considering only the $\pi\pi$ scattering.  
Conclusion about the $f_0 (1710)$ is quite consistent with the experimental  
facts that this state is observed in $\gamma\gamma\to K_S{\bar K}_S$~\cite{22}  
but not observed in $\gamma\gamma\to\pi^+\pi^-$~\cite{23}.  
  
As to the $f_0(1500)$ ($m_{res}=1510$ MeV, $\Gamma_{tot}=397$ MeV)  
we suppose that it is the eighth component of octet mixed  
with the glueball being dominant in this state. Its largest width  
among the enclosing states points also to its glueball nature  
\cite{24}. Note that in the PDG tables on the $f_0(1500)$ listing,  
an average value for the width of $109\pm7$~MeV is cited. However,  
there one indicates only the results of analyses of meson production  
processes, and in the few cases where the results of combined  
analyses of coupled processes are cited, authors did not use the  
representations of the multi-channel resonances by pole clusters  
(this is especially important in the case of wide resonances), i.e.,  
they did not apply all aspects of the multi-channel analysis.  
On the other hand, one can see from the data on scattering processes,  
analyzed here~\cite{Hya73}, that the energy dependence of observed  
quantities do not demonstrate a pronounced structure in the 1500~MeV  
region, which is needed for the narrow resonance. Therefore, it is  
reasonable to suggest that in this region there is a superposition  
of two states, a wide and a narrow one.  
  
It is known that there is a number of properties of the scalar mesons  
which do not allow for a satisfactory setup the lowest nonet.  
The main observations are the approximate equal masses of $f_0(980)$ and  
$a_0(980)$ and the $s{\bar s}$ dominance in the wave function of the  
$f_0(980)$. If these states are in the same nonet then the $f_0(980)$  
must be heavier than $a_0(980)$ by about 250-300 MeV due to the mass  
difference of $s$- and $u$-quark. Exclusion of the $f_0(980)$ as a non  
$q{\bar q}$ state and discovery of the $K_0^*$-doublet (if it will be  
confirmed) moves off a number of these problems.  
  
One can propose the following assignment of scalar mesons lying below  
1.9~GeV to lower nonets \cite{SBKN-prd10}. The lowest nonet: the  
isovector $a_0(980)$, the isodoublet $K_0^*(900)$, and $f_0(600)$ and  
$f_0(1370)$ as mixtures of the 8th component of octet and the SU(3)  
singlet. Then the Gell-Mann--Okubo (GMO) formula of  
\eq  
3m_{f_8}^2=4m_{K_0^*}^2-m_{a_0}^2  
\en  
gives $m_{f_8}=910$ MeV.  
For this nonet is seems to be important to test the nature of strange  
scalar meson $K_0^*(900)$ in a model-independent way. This will be  
the subject of a forthcoming paper~\cite{SBL}.  
  
In the relation for the masses of the nonet  
\eq\label{Rel_1}  
m_\sigma+m_{f_0(1370)}=2m_{K_0^*}  
\en  
the left-hand side is by about 18 \% larger than the right-hand one.  
  
For the next nonet {of the radial excitations} we find:  
$a_0(1450)$, $K_0^*(1450)$, and $f_0(1500)$ and $f_0(1710)$, the  
$f_0(1500)$ being mixed with a glueball which is dominant in this  
state. From the GMO formula we set ~~ $m_{f_8}\approx1453$ MeV.~ In the  
formula  
\eq\label{Rel_2}  
m_{f_0(1500)}+m_{f_0(1710)}=2m_{K_0^*(1450)}  
\en  
the left-hand side is by about 12.5 \% larger than the right-hand one.  
  
This assignment removes a number of prior questions and  
does not rise new ones. The mass formulas indicate to a  
non-trivial mixing scheme. Breaking of the relations  
(\ref{Rel_1}) and (\ref{Rel_2}) tells us  
that the $\sigma\!-\!f_0(1370)$ and $f_0(1500)\!-\!f_0(1710)$  
systems get additional contributions absent in the $K_0^*(900)$  
and $K_0^*(1450)$, respectively. A search of the adequate mixing  
scheme is complicated by the circumstance that here there is also  
a remainder of chiral symmetry, though, on the other hand, this  
permits one to predict correctly, e.g., the $\sigma$-meson mass.  
  
\begin{acknowledgments}  
  
The authors thank Thomas Gutsche and Mikhail Ivanov for useful  
discussions.  
This work was supported in part by the Heisenberg-Landau Program,  
the RFBR grant 10-02-00368-a, the Votruba-Blokhintsev Program for  
Cooperation of the Czech Republic with JINR (Dubna), the Grant Agency  
of the Czech Republic (Grant No.202/08/0984) and by Federal Targeted  
Program "Scientific and scientific-pedagogical personnel of  
innovative Russia" Contract No. 02.740.11.0238.  
  
\end{acknowledgments}

\appendix\section{Analytic continuation of  
the 3-channel $S$-matrix elements to unphysical sheets}  
  
Here we show, for convenience,  formulas of the analytic continuations of  
the 3--channel $S$--matrix elements to unphysical sheets of the Riemann  
surface in terms of those on sheet I (the physical sheet) --  
$S_{\alpha\beta}^I$ that  
have only zeros (beyond the real axis) corresponding to resonances, at least,  
around the physical region. In Ref.~\cite{KMS96} the  
general formula was given for the case of N channels and as example for  
three channels. The direct derivation of these formulas requires rather bulky  
algebra. It can be simplified if we use a circumstance that the $K$-matrix has  
the same value in all sheet of the Riemann surface of the $S$-matrix. This fact  
follows from Hermiticity of the $K$--matrix $K=K^+$ which means that the  
$K$--matrix does not try discontinuity when going across unitarity cuts. Then  
after some algebra, one can obtain formulas under interest shown below in the  
table.  
\begin{table}[htb]  
\caption{Analytic continuations of  
the 3-channel $S$-matrix elements to unphysical sheets}  
\label{tab:An_contin}  
\begin{center}  
\begin{ruledtabular}  
\begin{tabular}{ccccccccc}  
{Process} & I & II & III & IV & V & VI & VII & VIII \\  
\hline  
$1\to 1$ & $S_{11}$ & $1/S_{11}$ & $S_{22}/D_{33}$ & $D_{33}/S_{22}$ &  
$\det S/D_{11}$  
& $D_{11}/\det S$ & $S_{33}/D_{22}$ & $D_{22}/S_{33}$\\  
$ 1\to 2$ & $S_{12}$ & $iS_{12}/S_{11}$ & $-S_{12}/D_{33}$ & $iS_{12}/S_{22}$  
& $iD_{12}/D_{11}$ & $-D_{12}/\det S$ & $iD_{12}/D_{22}$ & $D_{12}/S_{33}$\\  
$2\to 2$ & $S_{22}$ & $D_{33}/S_{11}$ & $S_{11}/D_{33}$ & $1/S_{22}$ &  
$S_{33}/D_{11}$  
& $D_{22}/\det S$ & $\det S/D_{22}$ & $D_{11}/S_{33}$\\  
$1\to 3$ & $S_{13}$ & $iS_{13}/S_{11}$ & $-iD_{13}/D_{33}$ & $-D_{13}/S_{22}$  
& $-iD_{13}/D_{11}$ & $D_{13}/\det S$ & $-S_{13}/D_{22}$ & $iS_{13}/S_{33}$\\  
$2\to 3$ & $S_{23}$ & $D_{23}/S_{11}$ & $iD_{23}/D_{33}$ & $iS_{23}/S_{22}$  
& $-S_{23}/D_{11}$ & $-D_{23}/\det S$ & $iD_{23}/D_{22}$ & $iS_{23}/S_{33}$\\  
$3\to 3$ & $S_{33}$ & $D_{22}/S_{11}$ & $\det S/D_{33}$ & $D_{11}/S_{22}$  
& $S_{22}/D_{11}$ & $D_{33}/\det S$ & $S_{11}/D_{22}$ & $1/S_{33}$  
\end{tabular}  
\end{ruledtabular}  
\end{center}  
\end{table}  
In Table \ref{tab:An_contin}, the  
superscript $I$ is omitted to simplify the  
notation, $\det S$ is the determinant of the $3\times3$ $S$-matrix on sheet  
I, $D_{\alpha\beta}$ is the minor of the element $S_{\alpha\beta}$, that is,  
$D_{11}=S_{22}S_{33}-S_{23}^2$, $D_{22}=S_{11}S_{33}-S_{13}^2$, $D_{33}=  
S_{11}S_{22}-S_{12}^2$, $D_{12}=S_{12}S_{33}-S_{13}S_{23}$, $D_{23}=  
S_{11}S_{23}-S_{12}S_{13}$, etc.

\end{document}